\documentclass [12pt]{article}
\usepackage{latexsym}
\usepackage{amsfonts}
\usepackage{acronym}
\newcommand{\be}{\begin{equation}}
\newcommand{\ee}{\end{equation}}
\newcommand{\bea}{\begin{eqnarray}}
\newcommand{\eea}{\end{eqnarray}}
\makeatletter
\@addtoreset{equation}{section}

 \makeatother
\begin{document}
\normalsize
\title {
The Higgs Field As The Cheshire Cat \\And His Yang-Mills "Smiles"
}
\author
{{\bf L.~D.~Lantsman}\\
 Wissenschaftliche Gesellschaft, bei
 Judische Gemeinde  zu Rostock,\\Wilhelm-K$\ddot u$lz Platz,6,
 18055,\\ Rostock,Germany; \\ llantsman@freenet.de \\\\
{ $~~~~$\bf V.~N.~Pervushin} \\
  Bogoliubov Laboratory of Theoretical Physics, \\  Joint Institute for
Nuclear
Research, \\ 141980 Dubna, Russia;\\ pervush@thsun1.jinr.ru}
\medskip
\maketitle
\begin {abstract}
The well-known Bogomol'nyi-Prasad-Sommerfeld  (BPS) monopole is
considered in the limit of the infinite mass of the Higgs field as a basis of
the Yang-Mills field vacuum with the finite energy density. In this limit the
Higgs field disappears, but it leaves its trace as the BPS monopole transformed
into the Wu-Yang monopole obtained  in the pure Yang-Mills theory by  a
spontaneous scale symmetry breaking in the class of functions with the zero
value of the  topological charge.
The topological degeneration of the BPS monopole  manifests itself as 
Gribov copies of the covariant Coulomb gauge in the form of the time integral
of the Gauss constraint. We also show that, in the considered  theory,  there is a zero mode
of the Gauss constraint leading to an electric monopole and an additional mass
of $\eta'$-meson in QCD. The consequences of the monopole vacuum in the form of a rising potential and
topological confinement are studied in the framework of the perturbation
theory. An estimation of the vacuum expectation value of the square of the
magnetic tension is  given through the $\eta'$-meson mass, and arguments in
favour of the stability of the monopole vacuum are considered. We also discuss
why all these "smiles" of the Cheshire cat are kept by the Dirac fundamental
quantization, but not by the conventional Faddeev-Popov integral.
\end{abstract}
\newpage
%\tableofcontents
%\newpage
\section{Introduction.}
The nature of  the vacuum of the Yang-Mills (YM) theory
in the Minkowski space is an open problem at the present time. 
There were a lot of attempts to solve this problem. \par
A typical feature of  these attempts was the construction of the
nontrivial physical vacuum in the Minkowski space  on the basis of
nonzero values of vacuum expectations  coinciding with  statistical
averages (classical fields). As an  example of these attempts
we should like to point out the work \cite{Mat}
 stimulated by the asymptotic freedom
formula as a criterion for instability of the naive perturbation
theory \cite{Gr}. However, these attempts did not take  account of the
topological structure of vacuum.\par
This structure of the vacuum of the Yang-Mills
theory was discovered  in the Euclidean space \cite{Bel}, and it means
that there exist classical \it in-, out-\rm
 vacuum states corresponding to different topological indices
 $\vert n> $ with zero values of energy, and
 tunnel transitions $\vert n>\to \vert n+1>$
 occur between them. These  transitions are described by \it instantons\rm,
 i.e.  Yang-Mills fields with  fixed topological numbers
 $\nu= n_{ out}- n_{ in}$ satisfying the minimum of the Yang-Mills action.
 A defect of this vacuum is  non-physical status of a zero value of energy
in quantum theory.
However,  the topological degeneration of 
initial data for Yang-Mills fields
does not depend on the space where these fields are considered.
The initial data of any classical solution in the Minkowski space-time
are also topologically degenerated. It is worth to investigate topologically
degenerated vacuum solutions in  the Minkowski space-time
in the class of functions with physical values of finite energy densities.
 \par
 The present paper is devoted to just this investigation of the
 nontrivial topological vacuum of  the  Yang-Mills theory
 in  the Minkowski space-time.
 This vacuum is stipulated  by the fact that the homotopy groups of
all the 3-dimensional  paths (loops) on the $SU(2)$  group manifold are
 \it nontrivial \rm (p.325 in  \cite{Al.S.}):
\be
\label{top1}
\pi_3 (SU(2)) =\bf Z   .
\ee
 The  Yang-Mills vacuum should take account of this topology.
 We investigate the topological degeneration of the initial
 data using the well-known \it Bogomol'nyi-Prasad-Sommerfeld \rm (BPS)
 monopole as an example. This  monopole is a result of the spontaneous
break-down of the $SU(2)$ symmetry on the basis of  the classical Higgs  field
$\phi_0$ (i.e. the  $\lambda \phi^4$ theory) in the limit  $\lambda\to 0$.
This means that we consider \it the ideal Bose gas \rm of
scalar Higgs particles. This Bose gas is called \it the Bose
condensate\rm. The Higgs Bose condensate has a direct
analogy with  Bose condensate in the theory of superfluid helium \cite{N.N.}.
\par
Thus, there is the possibility  to construct the YM vacuum using
the well-known Bose condensate of free scalar particles in the limit
of their infinite mass  when these particles disappear from the spectrum
of elementary excitations of the theory leaving their "traces" in
the form of  monopoles. The study of these "traces" is  the aim
of the present paper.
\par
One of these "traces" is the topological degeneration of the
BPS monopole perturbation theory that
 manifests itself as  \it Gribov copies \rm of
the covariant Coulomb gauge treated as initial data of the Gauss constraint 
in the lowest order of the perturbation theory with the new monopole vacuum.
The \it Gribov copies \rm mean that there is a zero mode
of the Gauss law constraint  expressed through the \it global
variable \rm $N(t)$ that describes a  topological motion of  the Yang-Mills Bose
condensate as a whole system with the real momentum spectrum.
\par
We construct the generating functional for weak perturbation excitations
over this vacuum in the form of the Feynman path integral.\par
The paper  is organized as
follows:\par
Section 2 is devoted to a brief review of the problem of  vacuum
in  quantum-field  theories. We discuss
what the vacuum  is in  such theories and the ways of symmetry
break-down, which were very  fruitful in modern physics.\par
Further we  give  the general picture of the symmetry  break-down
and its connection with the  nonzero vacuum expectation value
$<0\vert\phi \vert 0>\not =0$ with
the example of the $\lambda \phi ^4$ theory. \par
In the conclusion of Section 2 we  show  that  the
Higgs sector of the Yang-Mills theory in the BPS limit $\lambda \to 0$ leads to
 nontrivial monopole
solutions of  the  equations of motion with a finite energy density
corresponding to the $SU(2)\to U(1)$ spontaneous break-down.
The Bogomol'nyi equation, defining the lowest level of the monopole energy,
determines  the direct connection between Yang-Mills and Higgs multiplets.
This will be a starting point for the construction of a consistent
 theory of
the Yang-Mills vacuum in  Section 4.\par
 Section 3 is devoted to
 the construction of the Dirac variables in the 
Yang-Mills theory in the form of solutions of the Gauss
 constraint-shell equation. This will be the base of all our
 further consideration. \par
The topological degeneration of the initial data is the subject of
Section 4. We  argue in favour of that the vacuum in the "old" instanton
approach is not the physical one. Instead,  we construct the
monopole $\Phi_i^{(0)}({\bf x})$ in the form of the
stationary Bose condensate with the topological number 0 and the nonzero
"magnetic" tension $B(\Phi_i^{(0)})$ corresponding this monopole.
 All this is a result
of the $SU(2)$ spontaneous break-down, describing  by  the classical equations
of the non-Abelian theory in the class of fields with the topological
number 0. These equations have  nontrivial solutions in the form of
 Wu-Yang monopoles: $\Phi_i^{(0)}({\bf x})$. Our construction of the
 YM vacuum is only a presentation of such solutions  as BPS monopoles
 in the theory with Higss fields in the limit of their infinite masses,
  but with the finite energy density; so that
 the BPS "magnetic" tension $B(\Phi_i^{(0)})$ (in this limit)
 has itself a crucial importance.
 We show that in the considered limit the  Gibbs expectation value
 $<B^2>$ (defined as the averaging $B^2$ over the volume)
 can be different from zero; it is a direct
 analogy  with the \it Meisner effect \rm in a superconductor.
 In the language of the group theory, it means the spontaneous break-down of
 the $U(1)$ group.
This, in turn, is a precondition for the correct consideration
of the $\eta '$-meson problem in QCD. A nonzero value of
$<B^2>$ allows  also us to regularize our theory by the introducing of an
infrared cut-off parameter $\epsilon(<B^2>)$ that plays the role of the
size of the BPS monopole.\par
The goal of Section 4 is to show the nature of the topological
degeneration of  the  monopole $\Phi_i^{(0)}({\bf x})$ as the Gribov
 ambiguity of the covariant Coulomb  gauge  (in the form of the
 time integral of the Gauss constraint). 
This topological  degeneration
 is defined by the non-perturbation factor
 $\exp(n \Phi_0^{(0)}({\bf x}))$, where $\Phi_0^{(0)}({\bf x})$ is a solution
 of the Gribov ambiguity equation that
 coincides with the Higgs field in the form of a  BPS monopole.\par
 As the covariant Coulomb gauge is the time integral of the Gauss law
 constraint, the Gribov ambiguity signals that there are zero
 modes of the Gauss constraint considered as the equation for the time
 component $A_0^c(t,{\bf x})$ of the Yang-Mills field.\par
The in the main new step in our investigation is the introduction of the
non-integer, continuous topological variable $N(t)$ for the definition of the zero mode
of the solution of the Gauss law constraint for
$A_0^c(t,{\bf x})$ in the form of the product
$\dot N(t)\Phi_0^{(0)c}({\bf x})$. This zero mode induces the
"electric" tension ("\it electric monopole\rm ") as a dynamic degree of freedom that
cannot  be removed by fixing of any gauge.
This "electric" tension,
 in turn, generates  the action of a free rotator
describing the global rotation of the Yang-Mills vacuum as a whole system.
The corresponding
Schr\"odinger equation for vacuum has  the real spectrum of momenta in
contrast with the instanton case.
The dependence of the rotator action  on the Gibbs expectation value
 $<B^2>$, which, in  turn, depends on the Higgs mass
 (through the Bogomol'nyi equation), confirms our suggestion about
the Yang-Mills vacuum as a Bose condensate.\par
The topic of Section 5 is a more detailed analysis of  zero  modes
of the covariant Coulomb gauge and the Yang-Mills (constraint-shell)
action; also we decompose the "electric" tension into the
transverse and the longitudinal parts with respect to  the constraint-shell
equation.\par
 Section 6 is devoted to the calculation of the instantaneous potential of
 the current-current interaction in the presence of the Wu-Yang monopole.
  Instead of the Coulomb potential in QED, the corresponding
   Yang-Mills Green function take the form of the sum of the
 two potentials: the Coulomb potential and the rising
 one; it is of great importance for the  analysis of the
 hadronization, in particular of the $\eta '$-meson problem.\par
The analysis of the  Feynman and FP path integrals is the subject of
Section 7.\par
The last two  Sections, 8 and 9, are devoted to the analysis of the
topological confinement and the quark
confinement in QCD as direct consequences of the average over
the topological degeneration. \par
The theory considered in Sections 8   allow us to contend that only the
colourless ("hadronic") states form a complete set of physical
states in QCD. We  prove that
the topological confinement leads to the quarks  confinement in  QCD, that
 the complete set of  hadronic  states
ensures that  QCD is an unitary theory.
 In Section 10 we estimate the value of
  the vacuum chromomagnetic field in QCD${}_{(3+1)}$.
\section { Vacuum as a result of spontaneous symmetry break-down.}
\subsection 
 {A physical vacuum as a Bose condensate.}
All quantum -field  theories are considered in the \it Hilbert-Fock  space
of second quantization \rm (see \S 7.3 in \cite{Logunov} and
\cite{Leonid}).
It is quite logical to begin our  consideration with a suitable abstract
 mathematical model (p.40 in \cite {Logunov}). So, let some \it algebra
with involution \rm
$U$ be given (the creation -
annihilation operators are  examples of an algebra of that sort).
We denote this algebra by $C^*$. One constructs the $C^*$ - homomorphism
$\pi$ of the $C^*$ - algebra
$U$ into the algebra $B({\cal H})$\it $~$ of all the linear restricted operators \rm
defined in the Hilbert space $\cal H$. The homomorphism $\pi$ is called
\it the representation \rm of the $C^*$-algebra $U$ in the Hilbert space $\cal H$.
The representation $ \pi$ is called the \it irreducible one\rm, if  every closed
 subspace in $\cal H$\rm,
invariant with respect to all the operators  $\pi (A)( A\in U )$, is
$\emptyset$ or the whole Hilbert space  $\cal H$. The vector $\Phi \in \cal H $
is called  the \it cyclic vector \rm for the representation $ \pi$, if all the
vectors of  the form $\pi (A) \Phi$, where $ A \in U
$, form a complete set (\it a linear shell \rm ) in $\cal H$.
This representation with the cyclic vector is called
 the \it cyclic one\rm.\par
If $\Phi$ is  a vector in  $\cal H $, then it generates the
\it positive functional \rm
\be
\label{pos}
F_\Phi =< \Phi ,\pi (A)  \Phi >
\ee
on $U$ (in  terms of the  probability theory it is the
\it mathematical expectation \rm  of the value $\pi (A)$  in  the state
$\Phi$). This functional is called  the \it vector  functional
associated with the representation  $\pi $ \it  and  the vector \rm $\Phi$.\par
In these terms the \it Gelfand-Naimark- Sigal \rm (GNS) construction of a vacuum
(p.42 in \cite {Logunov}) consists in the following:
one can determine  some (cyclic)
representation  $\pi_F$   of the algebra  $U$ in the given Hilbert space
with  the cyclic  vector $\Phi_F$  for the
given positive functional  $F$  such that
\be
\label {GNS}
F(A)= <\Phi_F, \pi_F (A) \Phi_F >.
\ee
The representation   $\pi_F$  is determined with
these conditions as unique \it  to within the unitary equivalence\rm.
This construction of the
cyclic vector $\Phi_F$ as the \it vacuum  vector \rm (or simply
 the \it vacuum\rm) allowed us
to write down the theory of second quantization.\par
 Let us consider the free (i.e. without an interaction)
theory of  one  particle (boson).
This boson has its fixed  integer spin $s$ and mass $m$
(therefore  also the fixed square of the Pauli-Lubanski
vector $W^2=m^2s(s+1)$, and
its mass-shell equation is $p^2= m^2$). As usual in  quantum -field
theories, we place our  particle in a  large enough closed box. Then,
according to the \it Hilbert-Schmidt theorem \rm (p.231 in \cite{Kolm}),
the momentum spectrum of the particle is \it discrete\rm.\par
Associating the \it ortho-normalized \rm vectors of
 some infinite-dimensional Hilbert space with the quantum numbers of
 the momentum and spin (helicity),
we obtain  the \it one-particle Hilbert \rm (\it Fock \rm )\it space \rm
(at the level of  special relativity), which we denote as
${\bf \Sigma}^{[m,s]}$.\par
Let  some algebra with involution (depending on the momentum $p$ and spin $s$)
be given in the form of \it the Bose commutation relations\rm $~$
of the operators $a(p), a^{\dagger}(p)$ (the formula (4.16) in \cite{Ryder}):
\be
\label{CCR}
[ a(p), a^{\dagger}(p)]= (2\pi) ^3 2 \omega_p \delta^3({\bf p}- {\bf p} ')~,
\ee
where $\omega_p$ is the frequency corresponding to the momentum $p$, and
\be
\label {CCR1}
[ a(p), a(p')]=[ a^{\dagger}(p), a^{\dagger}(p')]=0.
\ee
 We suppose that the operator $a^{\dagger}(p)$
acts on the vacuum  vector (which we shall denote sometimes as
$\vert 0>$) as
\be
\label {v.act}
a^{\dagger}(p)\vert 0>=\Phi(p)
\ee
for the eigenvalue $p$ of  the momentum operator.
All such vectors, depending on $p$ (and the value of helicity),
form a linear shell of our ${\cal H}= {\bf \Sigma}^{[m,s]}$.
It is said that the boson with  the momentum $p$ and  spin $s$
(at a fixed helicity) \it was created from the vacuum \rm
$\vert 0>$.\par
Operator $a(p)$ has a contrary action on  the vacuum $\vert 0>$.
This action is expressed as
\be
\label{op.act}
a(p)\vert 0>=0~,
\ee
and we say that the operator $a(p)$ \it annihilates \rm  the vacuum $\vert 0>$.\par
The  considered simple example shows us how one can introduce the vacuum into
some quantum-field theory. This is a purely mathematical
construction, and  this vacuum is called  the \it mathematical  one\rm.
But, to construct a consistent relativistic
quantum-field  theory, we should   impose some conditions on its vacuum.
\par
First of all, this is the condition \it that the vacuum exists and  is
unique \rm to within  a  phase factor.
The  mentioned phase factor should  preserve the  (unit) norm of the
 vacuum  vector. Thus, this
factor has the form $\exp(i\phi)$, and we already have some
\it degeneration \rm of  the vacuum with respect to  the Abelian $U(1)$ group.\par
Then we demand that our  vacuum  should be invariant with respect to
 \it pure Poincare translations \rm $U(a,1)$ (p.251 in\cite {Logunov}). This
leads, as a final result, to the conservation of the momentum-energy tensor
of the theory. We associate always \it the minimum value of
energy \rm with the vacuum vector $ \Phi_0\in \cal H$.
  We call the state with the  minimum value of  energy  the \it ground physical
  state\rm. Note that,in the
free quantum -field  theory, the spectrum of the momentum-energy operator $P$
 belongs to the set  $\bar V^+_\mu\bigcup \{0\}$, where $V^+_m =
\{p\in M: p^2\geq m^2,p^0>0\}$ at $m\geq 0$
and $M$ is the Minkowski space (the so-called \it strict condition of
spectrality\rm).
% The infinitesimal parameter $\mu$ is called
% \it the mass crack\rm, and it plays an important role for the construction
 %of the effective scattering theory.
 \par
 All the  states are considered \it as \rm (\it perturbation\rm ) \it excitations
 over the vacuum\rm. We shall utilize this fact in our present
work.\par
The condition that vacuum is invariant with respect to  pure  Poincare
 translations is satisfied when $P=0$. All the states with such $P$ are
invariant under pure Poincare translations. All the
unitary representations of this class, except for  the unit representation,
 the vacuum ($U(a,
\tilde \Lambda)\equiv 1$, where $\tilde \Lambda\in SL(2,C)$)
are \it infinite-dimensional \rm with respect to the 3-dimensional
$SO(3)$ rotations.
\par
We say that the \it cluster feature \rm (or \it the feature of the asymptotic
factorisation\rm) is fulfilled in the  physical Hilbert space $\cal H$
 if there exists the vector of unit norm $\Psi_0\in \cal H$ such that
\be
\label{cluster}
<\Phi, U(\lambda a,1)\Psi >~\to~ <\Phi,\Psi_0><\Psi_0,\Psi>,
\quad \lambda\to \infty~,
\ee
where $a$ is an arbitrary space-like vector in the Minkowski space $M$;
$\Phi,\Psi \in \cal H$.\par
It turns out that the condition that the vacuum exists and is unique
\it is equivalent to the cluster feature \rm (\ref{cluster}).\par
Let us consider (p.294 in \cite {Logunov}) some quantum field $\phi$ and let
us supply it with the  index $\kappa$ ($\phi^{(\kappa)}$) which defines the
type of this field (for example its spin). Thus, every  $\phi^{(\kappa)}$
 is a tensor or a spin-tensor  with a finite number of its
Lorentz components: $\phi_l^{(\kappa)}$($l=1,...,r_\kappa $) and with a
definite transformation features with respect to the eigen Lorentz
group $L_+^\uparrow$ or its covering $ SL(2,C)$.\par
In these terms we can construct a consistent relativistic
quantum field theory if  its vacuum is \it cyclic \rm in the following sense.
The set $D_0$ of  finite linear combinations of  (spin)-tensor fields
of the form  $\phi^{(\kappa_1)}_{l_1}(f_1)...$
$\phi^{(\kappa_n)}_{l_n}(f_n)\vert 0>$ is \it dense \rm in $\cal H$.\par
A very important role in quantum-field  theories plays the vacuum
expectation value of some quantum field $\phi$ over  the vacuum $<0\vert\phi
\vert 0>$. But this value is \it zero \rm at the level of  the algebra of
 creation-annihilation operators $a(p),a^{\dagger}(p')$
in the Fock-Hilbert space of  second quantization: because of the
relation (\ref{op.act}).\par
Now, with the help of  simple arguments, we shall show the connection of the
relation $<0\vert\phi \vert 0>=0$ with the question of (global)
symmetry and its break-down (see for example \S 5.3 in \cite{Cheng}).\par
Let $U$  be an element of the (global) \it unitary realized
\rm symmetry group  with respect to which the Hamiltonian $H_0$ of some
quantum -field  theory is invariant. Then we can write down this condition
 of invariance of the Hamiltonian $H_0$ as
\be
\label{inv}
U H_0 U^{\dagger} =H_0.
\ee
If the considered quantum field  theory is realized in the physical Hilbert
space $\cal H$, and we constructed already the irreducible
representation $\pi (U)=U'$ in $\cal H$, then  some transformation of  the
group $U$ induces  the corresponding  transformation on  $\cal H$:
\be
\label{transf}
U' \Phi =\Psi.
\ee
If $E_\Phi$ and $E_\Psi$ are the expectation values of the Hamiltonian
$H_0$ in the states $\Phi$ and $\Psi$, respectively, then we can rewrite the
condition of invariance of the Hamiltonian $H_0$ as
\be
\label{H.i}
E_\Phi= <\Phi\vert H_0\vert\Phi>=<\Psi\vert H_0\vert\Psi>=E_\Psi.
\ee
Thus, the symmetry of the Hamiltonian $H_0$ means
\it  the degeneration of eigenstates of  the energy operator
\rm corresponding to the
irreducible representation  of the symmetry group. However, the  relations
(\ref{transf}),(\ref{H.i}) stay implicit at our suggestion
that vacuum exists and is unique. Really, since the states
 $\Phi$ and $\Psi$ should be connected with the ground state $\vert 0>$ by
the relations (see (\ref{v.act})):
\be
\label{v.act1}
\Phi= \phi ^{\dagger} \vert 0>,\quad \Psi= \psi ^{\dagger}\vert 0>
\ee
and
\be
\label{U'}
U' \phi ^{\dagger}U^{\dagger '}= \psi ^{\dagger}~,
\ee
then the relation (\ref{transf}) is true if and only if
\be
\label{inv.vac}
U'\vert 0>=0.
\ee
If the condition (\ref{inv.vac}) is not  fulfilled, then the condition 
(\ref{H.i}) is also broken,
 and, together with this fact, \it the conclusion about the symmetry
of degenerated levels of energy is also  broken\rm. This situation is
called  the \it spontaneous break-down of the symmetry \rm $U$.\par
Thus, we can write down the condition of the spontaneous break-down of some
symmetry as
\be
\label{br.down}
U'\vert 0> \neq 0
\ee
(\it the vacuum becomes not  invariant with respect to the  group \rm $U$).\par
If $U=\exp (i \epsilon^a Q_a)$, where $\epsilon^a$ are  continuous
 group parameters (\it Euler angles \rm ), and $Q_a$ are  \it generators \rm (\it charges \rm)
of this group, then (\ref{br.down}) is equivalent to that charges $Q_a$ do not
annihilate the vacuum $\vert 0>$:
\be
\label{ch.vac}
Q_a\vert 0>\neq 0.
\ee
The statement equivalent to (\ref{br.down}),(\ref{ch.vac}) is that the  field operators
 $\phi_i$ (considered \it as degrees of freedom in the
Lagrangian  formalism\rm; one can interpret them as
 components of the \it group multiplet\rm)\it  have
 nonzero vacuum expectation values\rm:
\be
\label{exp.val}
<0\vert \phi_i\vert 0>\neq 0.
\ee
The symmetry transformation (\ref{U'}) is equivalent to
\be
\label{field transf}
\phi_i(x)\to \phi'_i(x)= \phi_i(x)+\delta \phi_i(x),
\ee
where
\be
\label{delta phi}
\delta \phi_i(x)=i \epsilon_a t^{aj}_i\phi_j(x),
\ee
and $t^a$ are the matrices of the adjoin representation of the Lie algebra
of  the group $U$.
The application of the Lie algebra and its
adjoin  representation guarantees  the fulfilment of the \it N\"other
theorem \rm and the existence of conserved currents in the theory.
\par
As it follows from the N\"other theory, the conserved charges have the form
\be
\label{char}
Q^a=\int d^3x J_0^a(x)
\ee
with
\be
\label{cur}
J_0^a = -i \frac {\delta {\cal L}}{\delta \partial^0  \phi_i}t^a_{ij}\phi^j
\ee
($\cal L$ is the Lagrangian  of the theory).\par
Then, it is easy to see that
\be
\label{commut}
[Q^a,\phi_i]=i t^a_{ij}\phi^j.
\ee
Thus, (\ref{ch.vac}) means that at least  some  matrix elements of
$<0\vert \phi_i\vert 0>$ are different from zero.\par
One can show (\S 10.3 B in \cite {Logunov}) that \it the break-down of the
global symmetry is accompanied with the appearance of  \it Goldstone
massless and spinless bosons \rm (\it the  Goldstone theorem\rm).\par
Note, and this remark has a great importance for our statement,
that the conservation of the charges in the N\"other theory means that the
Lagrangian is invariant under  transformations of a (global) symmetry.
The latter means, in turn, that we should define always \it
 the minima of the Lagrangian\rm: in  fact \it the minima of the potential
 \rm $V(\phi_i)$.
 As we noted above, the minimum state of the
 potential energy corresponds to the vacuum vector $\vert 0>$.
 It is obvious that  the potential  $V(\phi_i)$ has its (global) minimum at
 $\phi_i= <0\vert \phi_i\vert>\equiv a_i$. If  the potential $V$ is a function
  of  several quantum fields: $\phi_i, i=1,...,n$, we should
solve a set of  equations of the first order
to define its (global) minimum. These equations describe
the minimum surface for the potential $V$
(for example, it is the circle $\sigma^2+ \pi^2 =a^2$
in the $\sigma, \pi $ model with the Abelian $U(1)$ symmetry
described in the
monograph \cite{Cheng},p.p.147-149),
providing invariance of the
Lagrangian  with respect to the symmetry translations.
The vacuum vector $\vert 0>$ remains
invariant at such translations: according to (\ref{inv.vac}).
However, this degenerated construction is not steady in general;
since  the vacuum  $\vert 0>$ is unique (because of  the  GNS theorem),
the symmetry would be broken down inevitably.\par
If the potential of the considered theory has, for example, the two
minima differing in  sign, we should choose  one of them.\par
The next example is  the Lagrangian of  the simplest and nevertheless
 very important $\lambda \phi^4$ theory having  the form
(p.10 in \cite{Linde})
\be
\label{scalar}
{\cal L}=  \frac {1}{2}( \partial_\mu \phi)^2+\frac {m^2}{2}\phi^2- \frac {\lambda}{4}\phi^4.
\ee
The points of global minimum of this Lagrangian are
$\phi_0=\pm m/\sqrt {\lambda}$, and we choose the positive sign.
 Note also that if the initial value  of $\phi$ is
zero, the symmetry  will be broken down spontaneous by
during the time $\sim m^{-1}$. Note also, and it will be very important
for our further consideration, that the vacuum configuration
$\phi_0=m/\sqrt {\lambda}$ describes some sphere $S^2$ in
the field configuration space.\par
 Considered in the present work the $SU(2)$ group is one of  the
examples of \it local \rm gauge symmetries,
which differs from  global symmetries by a dependence
of the group parameters on  coordinates.
The general sketch of the proof remains the same as it was for the case of
global symmetries. The key point here is also the nonzero vacuum
expectation value $<0\vert \phi_i\vert 0>$.\par
It is naturally to ask: if $<0\vert \phi_i\vert 0>\neq 0$ in the case of
a symmetry break-down, what is the physical nature of this value?
Let us consider this question with an example of the most simple
$\lambda \phi^4$ theory. \par
First of all,  note that the expectation value of the number
of particles in  some physical state (in our consideration we are
interested in the ground state) is
\be
\label{number}
n=<a^{\dagger}(p)a(p)>~;
\ee
it is the direct consequence of the  CCR (\ref{CCR}),(\ref{CCR1}). \par
Let us consider now a \it semi-classical \rm system of   Higgs
scalar particles $\phi$ (this means that $n$ is very large according to the
Bohr-Sommerfeld theory (\S 48 in \cite {Landau3}) \it in a state of 
thermodynamic equilibrium \rm  with the temperature $T$
(p.78 in\cite {Linde}).\par
Since the scalar particles have no conserved charges, and the
creation-annihilation processes have the equal probabilities in the free dynamics
theory, \it the chemical potential  $ \mu$ of the Higgs scalar is zero,
\rm and we can write down the \it small \rm Bose distribution for the
Higgs scalar
theory as
\be
\label{Bose}
n_p= \frac {1}{\exp(p_o/T)-1},
\ee
where $p_0=\sqrt {{\bf p}^2+m^2}$ is the energy of a particle
with  the momentum $\bf p$ and  mass $m$.
As $T\to 0$,  $n_p\to 0$. If $T\neq 0$, all the
physical important quantities (thermodynamic potentials,
Green functions, etc.) in the considered system are determined by
the \it Gibbs expectation values\rm:
\be
\label{Green}
<\phi>=\frac {Tr [\exp (-H/T\phi)]}{Tr[\exp(-H/T)]}~,
\ee
where $H$ is the Hamiltonian of the considered system. \par
 We can consider, in the limit $\lambda \to 0$,
the \it ideal  Bose gas \rm of  Higgs particles, $\phi_0$, which we
shall call henceforth as \it the Bose-condensate\rm. The \it collective
motion \rm of the above ideal  Bose gas is described  with the help of the
\it  stationary \rm (at the  zero momentum) theory, alike the situation
with superfluid helium in the Bogoliubov theory \cite{N.N.,Pervush1}.\par
As a direct consequence of  the CCR (\ref{CCR}),(\ref{CCR1})(see p.63 in \cite{Linde}),
we write down for the  stationary Bose condensate $\phi_0$:
\be
\label{distrib}
n_p= (2\pi)^3 \phi_0^2 m \delta ({\bf p}).
\ee
Then we decompose the Higgs field, described by the Lagrangian (\ref{scalar}),
into the \it Bose condensate \rm(\it which we call  the \it classical
field\rm, since it is described by the Bohr-Sommerfeld-Gibbs
theory, i.e. statistic physics) and the \it perturbation excitations
over the Bose condensate\rm, which we identify with the scalar particles.
The alike way was utilized in the work \cite {Pervush1}. \par
Thus, \it the symmetry break-down parameter $<0\vert \phi_i\vert 0>$
coincides in our case with the value of the  classical field $\phi_0$\rm.\par
% Note that, in the limit $\lambda \to \infty$, \it the symmetry cannot
%be restored\rm. The standard thermodynamic calculation of the Curie
%point for this restoration (p.16 in \cite {10}) yields
%$$
%T_c =2m/ \sqrt {\lambda}\to \infty ~.
%\tag 2.22
%$$
We can draw the following very important conclusions from the said above.
 Firstly, the presence of  the Higgs Bose condensate is a principal sign of
 the symmetry break-down; secondly, the  vacuum, in the form of the Bose
 condensate, has \it a nontrivial structure \rm and can be considered
as a \it real physical vacuum\rm. It is true, since the 
Bose distribution of momenta, (\ref{Bose}), is real.
All this is correct not only for the $\lambda \phi^4$ theory, but also for
 the  Yang-Mills theory, the aim of our present investigation. 
\subsection{ Gauge Higgs effect.}
Our idea, basing onto the above consideration, is to construct the consistent
  Yang-Mills vacuum, using the Higgs Bose condensate in the
theory with monopoles \cite{Hooft, Polyakov, BPS} in the well-known 
\it Bogomol'nyi-Prasad-Sommerfeld \rm (BPS) \it  limit
 of the zero self-interacti-\linebreak on\rm: $\lambda \to 0$(at $m \to 0$),  in  the Higgs sector of the
YM action  (see \cite{Al.S.,Gold}):
\be
\label{YM L}
S=-\frac {1}{4 g^2} \int d^4x F_{\mu \nu}^b F_b^{\mu \nu }+
\frac {1}{2} \int d^4x (D_\mu\phi,D^\mu\phi
) -\frac {\lambda}{4} \int d^4x \left[(\phi^b)^2- \frac{m^2}{\lambda}\right]^2~,
\ee
where $D_\mu\phi=\partial^\mu\phi+g[A^{\mu },\phi]$ is the
covariant derivative, $g$ is the coupling constant.\par
We suppose that  the initial data of all the  fields are given to within stationary
gauge transformations, the  manifold of these transformations has  a
nontrivial structure of   3-dimensional paths in the group space of
the non-Abelian $SU(2)$ gauge group:
\be
\label{3-path}
\pi_3 (SU(2)) =\bf Z~,
\ee
where  $\bf Z$ is the group of integers: $n=0,\pm 1,.\pm 2,..$.
\par
In the case of  the $SU(2)$ gauge theory
the Yang-Mills fields $A^{\mu b}$ and Higgs fields $\phi^b$ take
theirs values in the Lie algebra  of the  $SU(2)$ group.\par
If we want to obtain the fields corresponding to the action
\it with finite values of energy\rm, we should demand that the
field $\phi ({\bf r})$ is finite as ${\bf r} \to \infty$ in 
the \it Bogomol'nyi- Prasad-Sommerfeld \rm (BPS) \it limit \rm  $\lambda\to 0$.
This means that $\phi^a$ should go to
the minimum of the potential $V$:
\be
\label{Higs.as}
\phi^{a \infty} ({\bf n})\in M_0, \quad {\bf n} =\frac {\bf r}{r},
\ee
where $M_0$ is the  manifold of the minimum of the potential $V$(\it the vacuum
manifold \rm ):
\be
\label{min}
M_0=\{\phi= a;\quad a^2=m^2/ \lambda\}
\ee
as ${\bf r} \to \infty$. Thus, $M_0$ consists of the points of the sphere
$S^2$ in the 3-dimensional space of  the $SU(2)$ gauge symmetry, which is
broken down spontaneously  to the $U(1)$ group.
 We can choose the vector $\vec \phi $ along the axis $z$ in
 the Cartesian coordinates:
\be
\label{phi}
\vec \phi =(0,0,m/\sqrt{\lambda})
\ee
as a configuration of the ground state. Thus, this vector remains
invariant with respect to rotations around the axis $z$ (the  $U(1)$
transformations).
\par
Note, however, that the choice (\ref{phi}) in the whole space is
\it topological trivial \rm. Really (\cite {Al.S.}, \S $\Phi$4), the gauge condition
$\phi _i=0, i=1,2;
\phi_3=\vert \vec \phi\vert$ is impossible for the nontrivial
topologies $n\neq 0$. For the field satisfying such condition
its asymptotic at the spatial infinity
is trivial: the solutions of the  form $\vec\phi^{\infty}
({\bf n})= {\bf V} ({\bf n})\vec \phi$, where $V ({\bf n})$ is a
continuous function of $\bf n$ with its values in the  $SU(2)$
group in the case of the Yang-Mills theory, are topologic equivalent to
$\vec \phi =(0,0,a)$.
On the other hand, ${\bf V} ({\bf n})$, considered as the map $\pi _2 (SU(2))$,
is equal to zero.
\par
 We should define the topological structure of the
manifold (\ref{min}). First of all, because of  the GNS construction,
$\nabla \phi^a=0$. In
the case of some discrete group  $G$,  $\phi^{a \infty}$ should be
constant, since it is a continuous function
(from the topological point of view, we deal  in this case
with the group $\pi_0$ \cite{Switz} of the
connection components, which is trivial in the case of a
connected manifold; the sphere $S^2:=\{{\bf n} =1$ as $ {\bf r} \to \infty \}$ is
namely such case).
In this case $\phi^{a \infty}$ has a trivial topology. \par
If  $dim ( M_0) \neq 0$, $ M_0$ has a nontrivial topology.
Therefore, the group of symmetry $G$ \linebreak \it should be continuous\rm.
One can show (p.p. 465-466 in \cite{Cheng}) that the covariant
derivatives $D_i\phi $, taking part in the action  (\ref{YM L}), decreases as $r^{-2}$;
thus  the integral (\ref{YM L}) is finite. This guarantees  nontrivial topological
features of the theory.\par
Issuing from the formula (\ref{min}), which defines the manifold $M_0$ of the
minimum of the potential $V$, and demanding that $\vec \phi({\bf r})$
goes to some value of  $ M_0$, we see that the sphere $S^2\simeq M_0$
maps into the sphere $S^2:=\{{\bf n} =1\}$ as
 $ {\bf r} \to \infty$.
This map has the nontrivial homotopy group of 2-dimensional loops:
\be
\label{top2}
\pi_2 S^2= \pi_3 (SU(2))=\pi_1(U(1))=\pi_1~S^1=\bf Z~.
\ee
Namely this nontrivial topology determines \it  magnetic charges \rm
connected  with the remaining $U(1)$ symmetry (these charges alone point to
some electromagnetic theory). The presence of  magnetic charges means
that there exist solutions of the motion equations
for  the action  (\ref{YM L}) \it in the class of
magnetic monopoles\rm, i.e. \it the stationary vacuum solutions at the spatial  infinity
corresponding to the quantum-field configuration of the
minimum energy \rm $E_{min}$ (according to our definition of the
vacuum as a ground state of the minimum energy).
We can  write down these monopole solutions.\par
For example, the Higgs isovector should be proportional to  $\bf n$
as $ {\bf r} \to \infty$: in the light of the above said  about the map
$S^2\simeq M_0 \to
S^2:=\{{\bf n} =1\}$ as  $ {\bf r} \to \infty$. Thus, its form should be
\be
\label{hedg}
\phi^a \sim \frac {x^a}{r}f(r,a)
\ee
as $ {\bf r} \to \infty~$; $f(r,a)$ is some continuous function which does
not change the topology (\ref{top2}).\par
This solution for $\phi^a$ appears for the first time in  the work \cite {Polyakov}
and it is called \it the hedgehog\rm. A good analysis
of hedgehogs is conducted in the monograph \cite {Linde}(p. 114-116).
One can show
(\S $\Phi$11 in \cite {Al.S.}) that there
 exists the  solution of the motion equations 
(\it regular in a finite spatial
volume\rm) 
\footnote{ The statement that these 
solutions are regular in a finite spatial
volume means that we should consider
the topology (\ref{top2}) and the manifold $M_0$,
(\ref{min}), also with account of this finite spatial
volume. If we wish to adapt our theory to the needs of
QCD(we shall see how to do this in Sections 8,9), the spatial
volume  determined by the typical hadronic
size, $\sim 1$ fm.($\sim 5$ GeV$^{-1}$), is quite sufficient for
our consideration.} in the form
\cite {Al.S.,Gold}
\be
\label{sc monopol}
\phi^a =
 \frac{ x^a}{gr} f_0^{BPS}(r)~,~~~~~~~~~~~~~~~~~~~~~~~~~~~~~
 f_0^{BPS}(r)=\left[
 \frac{1}{\epsilon\tanh(r/\epsilon)}-\frac{1}{r}\right]~,
\ee
\be
\label{YM monopol}
 A^a_i(t,\vec x)\equiv\Phi^{aBPS}_i(\vec x) =\epsilon_{iak}\frac{x^k}{gr^2}f^{BPS}_{1}(r),~~~~~~~~
  f^{BPS}_{1}= \left[1 -
 \frac{r}{\epsilon \sinh(r/\epsilon)}\right]
\ee
 obtained in the  BPS limit
\be
\label{lim}
\lambda\to 0,~~~~~~m\to 0:~~~~~~~~~~
~~~~~\frac{1}{\epsilon}\equiv\frac{gm}{\sqrt{\lambda}}\not =0~.
\ee
This solution satisfies the \it potentiality condition\rm: 
\be
\label{Bog}
{\bf B} =\pm D\vec \phi~,
\ee
where $\bf B$ is the magnetic tension in the theory (\ref{YM L}).
This equation (called \it  the Bogomol'nyi equation\rm) is obtained
by the evaluation
of the lowest bound of energy:
\be
\label{Emin}
E= 4\pi {\bf m }\frac {a}{g},~~~~~~~~~~~~\,\,
~~~~~a=\frac{m}{\sqrt{\lambda}}
\ee
 (where $\bf m$ is the magnetic charge), for the  monopole solutions.\par
The outlines of the proof of  the formula (\ref{Bog})
are the following (see \S $\Phi$11 in \cite {Al.S.}).
Following  to the  't Hooft-Polyakov model \cite {Hooft,Polyakov},
let us introduce the "electromagnetic tension"
as the scalar product
\be
\label{tens}
F_{\mu\nu}= <F_{\mu\nu}^a, \frac {\phi_a}{a} >.
\ee
The magnetic tension corresponding to this tensor is
\be
\label{H}
H^a= \frac {1}{2} \epsilon ^{ajk} <F_{jk}^b,\phi_b > a^{-1}.
\ee
We can write down the magnetic charge $\bf m$ as a \it stream of the magnetic
tension $\bf H$ through an infinite removed sphere \rm
(multiplied on $(4\pi)^{-1}$):
\be
\label{m}
{\bf m} =\frac {1}{4\pi} \int d{\bf S} {\bf H} =
\frac {1}{4\pi}\int d^3 x ~\partial _i \epsilon ^{ijk} <F_{jk}^b,\phi_b >a^{-1}.
\ee
Note also that
$$
\epsilon ^{ijk} \partial _i<F_{jk}^b,\phi_b > = \epsilon ^{ijk} \nabla _i <F_{jk}^b,\phi_b > =$$
$$ =\epsilon ^{ijk} (<\nabla _i F_{jk}^b,\phi_b >+ <F_{jk}^b,\nabla _i\phi_b>)= \epsilon ^{ijk}<F_{jk}^b,\nabla _i\phi_b>
$$
(we utilized here the fact that the usual derivative $\partial _i$
coincides with the covariant derivative $\nabla _i$ for the gauge
invariant value $<F_{jk}^b,\phi_b >$; we took  account also of the
Bianchi identity $\epsilon ^{ijk}\nabla _i F_{jk}^b =0$). Therefore,
\be
\label{m1}
{\bf m }=\frac {1}{8\pi}\int d^3 x ~
\epsilon ^{ijk}<F_{jk}^b,\nabla _i\phi_b>a^{-1}.
\ee
Then we consider the inequality
\be
\label{in}
\int  dx ~<c,b> ~\leq ~ \frac {1}{2} \int dx~(<c,c>+<b,b> )
\ee
which follows from the relation $\int <c-b,c-b>dx  \geq 0$;  the equality
is reached in the case $c(x)=b(x)$ only (here $c(x)$ and $b(x)$ take
theirs values in ${\bf R}^n$). Applying the inequality (\ref{in}) to the tensors
$\frac {1}{2g}\epsilon ^{ijk}F_{jk}^b$ and $\nabla _i\phi_b$, we
obtain that
\be
\label{est}
\int d^3x~\frac{\epsilon^{ijk}}{2g}<F_{jk}^b,\nabla _i\phi_b> ~ \leq ~
\frac {1}{2}\int d^3x\{ \frac {1}{4g^2}<F_{jk}^b,F_{jk}^b> +
<\nabla _i\phi_b,\nabla _i\phi_b>\}.
\ee
The integral in  the right-hand side of (\ref{est})differs
from the energy $E$ of the configuration $(\phi, A_\mu^a)$ in the absence
of the potential term: 
\be
\label{self}
E_1= \frac {1}{4}\lambda \int d^3x~ [\phi^2 - a^2],
\ee
only. Since the left-hand side of (\ref{self}) differs in the factor only
from the magnetic charge, we
obtain  the estimation
\be
\label{est1}
{\bf m} \leq \frac {g}{4\pi a} (E- E_1).
\ee
In other words,
\be
\label{h.neq}
E\geq 4\pi{\bf m} \frac {a}{g} +E_1,
\ee
and the equality  is reached in the case
\be
\label{Bog1}
\frac {1}{g}\epsilon ^{ijk}F_{jk}^b =\nabla ^i\phi^b.
\ee
But this is the Bogomol'nyi equation (\ref{Bog}) written down in the index form (at the plus
sign).\par
Since $E_1\geq 0$, and the magnetic charge $\bf m$ takes the integers  only,
the energy $E$ of the configuration $(\phi, A_\mu^a)$ allows the estimation
\be
\label{est3}
E\geq 4\pi\frac {a}{g}
\ee
in the general case.\par
In the BPS limit $\lambda \to 0$, if other parameters $(a,g)$ remain invariable,
this estimation becomes exact:
\be
\label{energy}
E= \frac {1}{4g^2} \int d^3x~ <F_{jk}^b,F_{jk}^b>
 +\frac {1}{2}\int d^3x~ <\nabla _i\phi_b,\nabla _i\phi^b>.
\ee
If the fields $(\phi, A_\mu^a)$ satisfy the  Bogomol'nyi equations (\ref{Bog}),(\ref{Bog1}),
the functional  (\ref{energy}) \linebreak reaches its minimum (\ref{Emin}).\par
The solutions of the Bogomol'nyi equation (\ref{Bog}) are 
definite formulas for the Yang-Mills fields
$A_\mu^a$ and   Higgs multiplet $\phi^a$
(depending only on $\bf x $) \cite {Al.S.,Gold}.
 The resembling  formulas will appear in our work on the basis of a
similar theory. We shall discuss them in Section 4.\par
Thus, we see that the Bogomol'nyi equation (\ref{Bog}) allows us to obtain a
consistent theory involving  the Yang-Mills and Higgs multiplets and
yielding the solutions of the BPS monopole type.
The Higgs sector of that theory defines the $U(1)$ group of symmetry
with a nontrivial
topology, i.e. with  magnetic charges and with  radial magnetic fields.
All this can be a  base for the construction of a similar theory
for the Yang-Mills vacuum in the Minkowski space.\par
In contrast to the "old" approach to  the Yang-Mills vacuum,
our  conception of the  Yang-Mills vacuum as a stationary Bose condensate
yields the
real spectrum of momentum. We shall  show  that the stationary vacuum
fields have the winding number $n=0$ and
they have the form of BPS ( Wu-Yang) monopoles.
The "electric" and "magnetic" tensions  corresponding to these vacuum fields
will also be constructed.
The topological degeneration ($n\neq 0$) in our theory is
realized due to  \it Gribov copies \rm of  the covariant Coulomb
gauge imposed
on the vacuum potentials. The Yang-Mills (gluon) fields are
considered as  weak perturbation excitations (\it multipoles\rm )
over this vacuum.
These
excitations  have the asymptotic $O(\frac {1}{r^{1+l}}),l>1$ at the
 spatial infinity.
\section{ The Dirac quantization of the Yang-Mills theory.}
Let us consider the  Yang-Mills theory with the  local $SU(2)$ group
in the four-\linebreak dimensional Minkowski space-time. The action of this theory
is given by the formula
\be
\label{act YM}
W[ A _\mu]=-\frac {1}{4} \int d^4x F_{\mu \nu}^a F_a^{\mu \nu } 
 = \frac {1}{2}\int d^4x (F_{oi}^{a2}- B_i^{a2})~,\ee
where the standard definitions of the non-Abelian "electric" tension $F_{oi}^{a}$:
\be
\label{electr}
F_{0i}^a = \partial _0 A^a_i- D(A)^{ab}_i A_{0b}, \quad  D^{ab}_i = (\delta ^{ab}\partial_i+ g \epsilon ^{acb}A_{ci})~,
\ee
and the "magnetic" one, $B_i^a$:
\be
\label{magnet}
B_i^a= \epsilon_{ijk} (\partial^j A^{ak} +\frac {g}{2}\epsilon ^{abc}A_{b}^j A_{c}^k)~,
\ee
are used. The action (\ref{act YM}) is invariant with respect to the gauge transformations $u(t; {\bf x})$:
\be
\label{gauge}
{\hat A}^u_i=u(t; {\bf x})({\hat A}_i+\partial_i)u^{-1}(t,{\bf x}),\quad \psi^u :=u(t; {\bf x})\psi ,
\ee
where ${\hat A}_\mu= g \frac {\tau ^a}{2i}A_{a\mu} $, and $\psi^u $ is a spinor field.\par
 Solutions of the non-Abelian constraint equation
(\it  the Gauss law constraint \rm ):
\be
\label{Gauss}
\frac {\delta W}{\delta A^a_0}=0,\Longleftrightarrow [D^2(A)]^{ac}A_{0c}= D^{ac}_i(A)\partial_0 A_{c}^i~,
\ee
and the motion equation:
\be
\label{mot}
\frac {\delta W}{\delta A^a_i}=0,\Longleftrightarrow [\delta_{ij}D^2_k(A)-D_j(A)D_i(A)]^{ac}A_{c}^j= D^{ac}_0(A)[\partial^0 A_{ci}-
D(A)_{cbi} A^{ob}]~,
\ee
are determined by  boundary conditions and  initial data.
They  generalize  the corresponding equations
in the  Maxwell electrodynamics
(see the formulas (7),(8) in \cite{Pervush2}). \par
 The Gauss law constraint  (\ref{Gauss}) connects  the initial data of $A^0_a$
 with the one of  the spatial components $A^i_a$.
To remove the non-physical variables, we can solve this constraint in the form of the naive perturbation
series:
\be
\label{ser}
A^0_c= \frac {1}{\Delta}\partial_0 \partial_iA_{c}^i +... ,
\ee
where $\Delta$ is the Laplacian. As we remember from mathematical physics (see for example p.203 in
\cite {Vlad}),
the \it fundamental
solution \rm
of the \it Laplace equation\rm:
\be
\label{Laplace}
\Delta {\cal E}_3 =\delta (x),
\ee
is
\be
\label{fund}
{\cal E}_3 = -\frac {1}{4\pi x}.
\ee
This defines the action of the operator $\Delta^{-1}$ on
some continuous function $f(x)$:
\be
\label{Delta}
\Delta^{-1}f(x)=- \frac {1}{4\pi } \int d^3 y\frac {f(y)}{\vert x-y\vert},
\ee
that is  the \it Coulomb kernel of the non-local distribution \rm
(see also (12) in \cite {Pervush2}).\par
Thus, the resolving of the constraint and the substitution of this
solution into the equations of motion distinguishes  the
gauge-invariant non-local (radiation) variables. After the substitution of this solution into the equation
(\ref{mot}) the lowest order of this equation in the
coupling constant $g$ contains \it only  transverse  fields \rm
(this level coincides mathematically, as a \it linearized Yang-Mills
theory\rm, with the theory of  radiation variables in QED\cite {Pervush2}):
\be
\label{QED}
[\partial_0^2-\Delta] A^{cT}_k+...=0 , \quad A^{cT}_i= [\delta_{ik}- \partial_i\Delta^{-1}\partial_k]A^{ck} +...
\ee
This perturbation theory is well-known as  the \it radiation\rm \cite{Schwinger}
or \it  Coulomb \rm  \cite{Fadd1,Gitman} gauge
with the generating functional of Green functions in the
form of a Feynman integral in the rest
frame of reference $l^{(0)}=(1,0,0,0)$:
\bea
\label{Fein}
Z_F[l^{(0)},J^{aT}]&=&\int\limits_{ }^{ }
 \int\limits_{ }^{ }\prod\limits_{c=1 }^{c=3 }
 [d^2A^{cT} d^2 E_{cT}]\nonumber\\
 &&\times\exp\left\{iW^T_{l^{(0)}} [A^T,E^T]-i\int\limits_{ }^{
 }d^4x [J^{cT}_{k} \cdot A_{cT}^{k}]\right\}~,
\eea 
with the constraint-shell action:
\be
\label{csha}
W_{l^(0)}^T[A^T,E^T] = W^I\vert \sb {\frac{\delta W^I}{\delta A_0} =0}~,
\ee
given in the  first order formalism (\cite{Slavnov},p.83):
\be
\label{WI}
W^I = \int d t\int d^3 x \{ F^c_{0i}E_c^i- \frac {1}{2}[E^c_i E^i_c+ B^c_i B^i_c]\}=\int d t\int d^3 x (E^{ci}\partial_0 A_{ci}+A_{0c}D^c-
H)~,
\ee
where
\be
\label{der}
D^c=\partial_k E^{kc}-g[A_k^b,E^{kd}]\epsilon _{bd}^c~,
\ee
and
\be
\label{Ham}
H=\frac {1}{2} (E_k^{c2}+B_k^{c2})=\frac {1}{2}[(E^{Tc})^2+(\partial_i \sigma ^c)^2+B_k^{c2}]
\ee
is the Hamiltonian of the Yang-Mills theory.
We decompose here the "electric tensi-\linebreak on"$ E^{kc}$ \it into the
transverse and longitudinal parts\rm:
\be
\label{decomp}
E^c_i= E^{Tc}_i+ \partial_i \sigma ^c, \quad \partial_i E^{Tc}_i= 0 .
\ee
The constraint
\be
\label{constr1}
\frac{\delta W^I}{\delta A_0}=0 \Longleftrightarrow D^{cd}_i (A)E_d^i=0
\ee
can be solved in terms of these (\it radiation\rm) variables.
The function $\sigma ^a$ has the form \cite {Fadd1}
\be
\label{sigma}
\sigma ^a[A^T,E^T]= (\frac {1}{D_i (A)\partial^i})^{ac}\epsilon_{cbd}A_k^{Tb}E^{Tkd}\equiv (\frac {1}{\hat \Delta})^{ac}
\epsilon_{cbd}A_k^{Tb}E^{Tkd}.
\ee
Note (see (16.24) in \cite {Gitman}) that  $Det$ $[D_i(A)\partial^i]$ in (\ref{sigma})
is \it the Faddeev-Popov \rm (FP) \it determinant \rm in the YM Hamiltonian formalism:
\be
\label{FP}
\hat \Delta ^b_a A_{0b}-\partial_i E^i_a=0,\quad \hat \Delta ^b_a \equiv D ^b_{ai}\partial^i,
\ee
where
\be
\label{moment}
E_{ia}= \frac {\partial {\cal L}}{\partial \dot A^{ia}}=F_{0i}^a
\ee
is the canonical momentum (\ref{electr}).\par
A complete proof that $det$ $\hat \Delta ^b_a$ is the FP determinant of the YM theory
is given in  the monograph \cite {Gitman}, where  it was shown that the  radiation gauge
in the YM theory is equivalent to the FP determinant
$det$ $\hat\Delta ^b_a$ (see (16.30)in \cite {Gitman}).\par
The operator quantization of  the Yang-Mills theory in terms of the
radiation variables belongs to Schwinger \cite {Schwinger},
who proved the relativistic covariance of  the radiation variables (\ref{QED}).
This means that the radiation  fields are transformed as
 non-local functionals (\it Dirac  variables\rm\cite {Pervush2}),
\be
\label{Dir.v}
{\hat A}_k^T[A]= v^T[A]({\hat A}_k+\partial_k) (v^T[A])^{-1},\quad {\hat A}_k^T = 
g\frac {A_k^{Ta}\tau _a}{2i},
\ee
where the matrix $v^T[A]$ is defined from  the condition of
transversality: $\partial_k A^{kT}=0$.
At the level of the Feynman integral, as we have seen in QED,
the relativistic covariance means the relativistic transformation of sources
(this led to the change of  the variables (29) in the  work \cite {Pervush2}).\par
The definition (\ref{Dir.v}) can be interpreted as
a transition to  new variables, allowing  us to
rewrite the Feynman integral in the form of the FP integral \cite{Fadd1,Fadd2,Fradkin1}:
\bea
\label{ymfpi}
 Z_F[l^{(0)},J^{aT}]&=&\int\limits_{ }^{ }
 \int\limits_{ }^{ }\prod\limits_{c=1 }^{c=3 }
 [d^4A^{c} ]\delta(\partial_i A^{ci} )Det[D_i(A)\partial^i] \nonumber\\
 &&\times\exp\left\{i W [A]-i\int\limits_{ }^{
 }d^4x (J^{T}_{ck} \cdot A^{Tkc}[A])\right\}~.
 \eea
It was proved in \cite {Fadd1,Fadd2,Fradkin1} that, \it on  mass-shells of
 radiation  fields, the scattering amplitudes do not depend on the factor
\rm $ v^T[A]$. But the following question is quite reasonable:
\it why we can not
 observe these scattering amplitudes\rm?
There are a few answers to this question: the infrared instability of the
 naive perturbation theory \cite{Mat,Nils},
the Gribov ambiguity, or the zero value of the FP determinant \cite {Gribov},
the topological degeneration of the physical states \cite{Pervush3,Ilieva,Nguyen}.
\section{Topological degeneration of initial data.}
\subsection{Instanton theory.}
One can  find a lot of solutions of equations of classical electrodynamics.
 The nature chooses the two types of  functions:
the \it monopole \rm (the electric charge) that determines
non-local electrostatic phenomena (including instantaneous bound states)
 and
the \it multipoles \rm that determine the
spatial components of  gauge  fields with the nonzero magnetic tensions.\par
The spatial components of  non-Abelian  fields, considered above as the
radiation variables (\ref{QED}) in the naive perturbation theory
(\ref{ser}), are also defined as multipoles.
In the non-Abelian theory, however, it is a reason, as we saw this in
Section2, to assume that \it
the spatial components of  non-Abelian  fields belong to the
monopole class of functions \rm like the time components of the Abelian
fields (as  the Coulomb potential  for example).
\par
This fact was revealed by the authors of  the instanton theory \cite {Bel}.
Instantons satisfy the duality equation in the Euclidean space
(where the Hodge duality operator $*$ has the $\pm 1$ eigenvalues for
 external 2- forms defining the Yang-Mills tension tensor);
thus, the instanton action coincides with  the \it Chern-Simons functional
\rm
( the \it Pontryagin index\rm) (see the formula (10.104) in \cite {Ryder}):
\be
\label{Ch-S}
\nu [A]=\frac {g^2}{16\pi ^2}\int\limits_{t_{in}}^{t_{out}}
dt\int d^3 x F_{\mu \nu}^a {{}^*F}^{\mu \nu}_a = X[A_{out}]-X[A_{in}]=
n(t_{out})-n(t_{in}),
\ee
where ( (10.93) in \cite {Ryder})
\be
\label{wind}
X[A]=-\frac {1}{8\pi ^2}\int{\sb V} d^3 x \epsilon ^{ijk}
Tr [{\hat A}_i \partial_j{\hat A}_k- \frac {2}{3}{\hat A}_i{\hat A}_j
{\hat A}_k],\quad A_{in,out}= A(t_{in,out},x)
\ee
is the \it topological winding number functional of  gauge  fields\rm,
 and $n $ is the value of this functional for the classical vacuum:
\be
\label{cl.vac}
{\hat A}_i= L^n_i= v^{(n)}({\bf x})\partial_i v^{(n)}({\bf x})^{-1}.
\ee
The manifold of all the classical vacua  in a non-Abelian theory
represents \it the group of  three-dimensional paths \rm lying in the
three-dimensional $SU(2)$-manifold with the   homotopy group
$\pi_3 (SU(2))= {\bf Z}$.
The whole group of stationary matrices is split into the
topological classes marked by the  integer numbers
( the \it degrees of the map\rm)
defined by the expression ((10.106) in \cite {Ryder}) 
\be
\label{degree}
{\cal N}[n]=- \frac {1}{24\pi ^2}\int d^3 x \epsilon ^{ijk}Tr [L^n_iL^n_jL^n_k],
\ee
which shows how many times the three-dimensional path $v({\bf x})$
 turns around the $SU(2)$-manifold when the co-ordinate $x_i$
runs over the space where it is defined.\par
Gribov, in 1976, proposed to consider the  instantons as
\it  Euclidean solutions interpolating between the classical vacua with different
degrees of the map (\it as tunnel transitions between these classical vacua\rm).\par
The degree of the map (\ref{degree}) can be considered as a condition for 
normalization that determines the class of functions with given
classical vacua (\ref{cl.vac}). In particular, to obtain the equation (\ref{cl.vac}),
we should choose the classical vacuum in the form
\be
\label{v}
v^{(n)}({\bf x})=\exp (n {\hat \Phi}_0({\bf x})),
\quad {\hat \Phi}_0 =- i\pi \frac {\tau ^a x_a}{r} f_0 (r)\quad (r= \vert {\bf r}
\vert )
\ee
(compare with (16.34) in \cite{Cheng}; we should set $x_0=0$ in this formula
for the stationary gauge transformations which we discuss now).
The function $f_0 (r)$ satisfies the boundary conditions 
\be
\label{cond}
f_0 (0)=0,\quad f_0 (\infty)=1.
\ee
Note a direct parallel between this solution and the formula (\ref{hedg}).
The common between the monopole and instantons theories is \it
the same nature of topologies\rm.
In the case of  the Yang-Mills  instanton theory we deal  with the map (\ref{3-path}):
$S^3\to SU(2)$ as
${\bf x} \to \infty$. This induces the homotopy group
$\pi_3(SU (2))= \pi_3 S^3 ={\bf Z}$ ($ S^3$ is the bound of the Euclidian space
$E_4$)
coinciding with $\pi_2 S^2=\bf Z$ (see (\ref{top2}))
in the theory (\ref{YM L})-(\ref{Bog}).
This generates similar theories. But there exists also the principal
distinction of  the both theories. As a consequence of the  relation 
$\pi_3(SU (2))= \pi_3 S^3 ={\bf Z}$, the 
instantons \it can exist in the YM theory
without any spontaneous $SU(2)$ break-down\rm. 
This break-down is not the necessary thing in this case,
and
we consider $SU(2)$ as an \it exact symmetry \rm in the instanton
theory.\par
 Thus, we obtain the  solution of the monopole type
in (\ref{v}) as ${\bf x} \to \infty$.\par
To show that these classical values are not sufficient
to describe  physical vacuum in the non-Abelian theory, we consider
the \it quantum instanton\rm, i.e. the corresponding zero vacuum solution of
the Schr\"odinger equation
\be
\label{Schrod}
{\hat H} \Psi _0[A]=0 ~,
\ee
where
${\hat H}= \int d^3x [E^2+B^2], E= \frac {\delta}{i\delta A }$
are operators of the Hamiltonian and field momentum respectively.
This solution can be constructed by using the winding number functional (\ref{wind})
and its derivative,
\be
\label{d.wind}
\frac {\delta}{\delta A^c_i} X[A]=\frac {g^2}{16\pi ^2} B^c_i(A).
\ee
The vacuum wave functional, in terms of the winding number functional
(\ref{wind}), has the form \it of a  plane wave  \rm \cite {Pervush1}:
\be
\label{plan}
\Psi _0[A]=\exp (iP_{\cal N}X[A])
\ee
for \it non-physical \rm  imaginary values of the topological
momentum $P_{\cal N}=\pm 8\pi i/g^2$ \cite {Pervush1,Arsen}.
We would like to note that in QED
 this type of the wave functional belongs to the non-physical part of the
spectrum like the wave function of the oscillator
 $({\hat p}^2+q^2)\phi_0=0$.
The value of this non-physical plane wave functional
\footnote{The wave function (\ref {plan}) is not normalized, the
imaginary 
topological
momentum $P_{\cal N}=\pm 8\pi i/g^2$ turns it in a
function with the non-integrable square.}
for the  classical
vacuum (\ref{cl.vac}) coincides with the \it semi-classical instanton wave function
\rm
\be
\label{wave}
\exp (iW [A_{instanton}]=\Psi _0[A=L_{out}]\times \Psi^* _0[A=L_{in}]=
\exp(-\frac {8\pi ^2}{g^2}[n_{out}-n_{in}]).
\ee
This exact relation between the semi-classical instanton and its quantum
version (\ref{Schrod}) points out that  classical instantons are also
non-physical solutions, they  tunnel permanently in the
Euclidean space-time between the classical vacua with  zero energies
that do not belong to the physical spectrum.
\subsection{Physical vacuum and  gauge Higgs effect.}
Our next  step is the assertion \cite{Jack} about the topological
degeneration of  initial data not only  of  the classical vacuum but also of
all the physical  fields with respect to the stationary gauge transformations
\be
\label{stat}
{\hat A}_i^{(n)}(t_0,{\bf x})= v^{(n)}({\bf x}){\hat A}_i^{(0)}(t_0,{\bf x})v^{(n)}({\bf x})^{-1}+L^n_i,
\quad L^n_i=
v^{(n)}({\bf x}) \partial _i v^{(n)}({\bf x})^{-1}.
\ee
The stationary transformations $ v^{(n)}({\bf x})$ with $n = 0$ are
 called \it the small one\rm; and those with $n\neq0$, \it the
large ones \rm  \cite {Jack}.\par
The group of transformations (\ref{stat}) means that \it
the spatial components of the non-Abelian  fields with  nonzero magnetic
 tensions \rm $B(A)\neq 0$ belong to the monopole class of
functions \rm  like the time components of the Abelian  fields.
In this case the non-Abelian  fields with  nonzero magnetic tensions
contain the non-perturbation monopole-type term, and the spatial components
can be decomposed into  sums of 
vacuum monopoles
$\Phi _i ^{(0)}({\bf x})$ and  multipoles ${\bar A} _i$:
\be
\label{sum}
A_i ^{(0)}(t_0,{\bf x})=\Phi _i ^{(0)}({\bf x})+ {\bar {\hat A}}_i ^{(0)}(t_0,{\bf x}).
\ee
Each multipole is considered as a weak perturbation part with the following 
asymptotic at the  spatial infinity:
\be
\label{asym}
{\bar  A}_i (t_0,{\bf x})\vert _{assymptotic}= O (\frac {1}{r^{l+1}})\quad (l>1).
\ee
Nielsen and Olesen \cite {Nils} and Matinyan and Savidy \cite {Mat}
introduced the  vacuum magnetic tension, using the fact that all the
asymptotically free theories are instable, and the perturbation vacuum is
not the lowest stable state.\par
The extension of the topological classification of  classical vacua to all the
initial data of the spatial components helps us to choose the
vacuum monopole with the zero value of the winding number functional (\ref{Ch-S}):
\be
\label{choose}
X[A=\Phi_i^{c(0)}]=0,\quad \frac {\delta X[A]}{\delta A_i ^c} \vert _{A=\Phi^{(0)}} \neq 0.
\ee
The zero value of the winding number, transverseness and spherical
symmetry (as a monopole) fix the class of initial data for  spatial
components:
\be
\label{vec}
{\hat\Phi}_i =-i \frac {\tau ^a}{2}\epsilon _{iak}\frac {x^k}{r^2}f(r).
\ee
They contain only one function $f(r)$. The classical equation for this
function has the form
\be
\label{eq}
D^{ab}_k(\Phi_i)F^{bk}_a(\Phi_i)=0 \Longrightarrow \frac {d^2f}{d r^2}+\frac {f (f^2-1)}{r^2}=0.
\ee
We can see the three solutions of this equation:
\be
\label{3}
f_1^{PT}=0,\quad f_1^{WY}=\pm 1\quad (r\neq 0).
\ee
The  first solution corresponds \it to the naive instable perturbation
theory with the asymptotic freedom formula\rm.\par
The two nontrivial solutions are well-known.
They are \it the Wu-Yang monopoles\rm, applied for the construction of physical variables
in the work \cite{Niemi} (\it the hedgehog and the antihedgehog \rm in
terminology of \cite{Linde,Polyakov}). As it was shown in the paper \cite{David1},
the Wu-Yang monopole leads to the rising potential of the instantaneous interaction with the
quasi-particle current. This interaction
rearranges the perturbation series, leads to the gluon constituent mass and removes the asymptotic freedom
formula \cite{Bogolubskaja,Yura}
as an origin of   instability.\par
The Wu-Yang monopole \cite{Wu} is a solution of the classical equations everywhere
\it besides the origin of coordinates\rm, $r = 0$. The
corresponding magnetic field is
\be
\label{B}
B_i ^a(\Phi_k)= \frac {x^ax^i}{gr^4}.
\ee
To remove the singularity at  the origin of coordinates and regularize its
energy, the Wu-Yang monopole is considered as a limit of
the \it Bogomol'nyi-Prasad-Sommerfeld \rm (BPS) monopole (\ref{YM monopol}):
\be
\label{ansatz}
f_1^{BPS}=[1-
\frac {r}{\epsilon \sinh  {r/\epsilon}}]\Longrightarrow f_1^{WY} ,
\ee
when the mass of the Higgs field goes to infinity in the limit 
of the infinite volume $V$:
\be
\label{mass}
 \frac{1}{\epsilon}=\frac{gm}{\sqrt{\lambda}}=\frac{g^2<B^2>V}{4\pi}\to \infty~.
\ee
The BPS monopole has the finite energy density:
\be
\label{magn.e}
\int \limits_{\epsilon}^{\infty } d^3x [B_i ^a(\Phi_k)]^2 \equiv V <B^2> 
= 4\pi \frac{gm}{g^2\sqrt{\lambda}}=\frac {4\pi} {g^2\epsilon} 
\equiv\frac{1}{\alpha_s\epsilon} ~.
\ee
(see also \cite{David2}).
The infra-red cut-off parameter  $\epsilon$ 
disappears in the limit $V \to \infty$, i.e. when the 
mass of the Higgs field goes to infinity and the Wu-Yang monopole
turns, in a continuous way, into the BPS monopole.  
In this case the BPS-regularization of the Wu-Yang monopole
is similar to the \it infrared regularization in QED
\rm by the introduction of
the "\it photon mass  \rm " $\lambda$ (see for example \cite {A.I.}, p.413) that also violates the initial
equations of motion
\footnote {It is necessary to note here that  in the
quantum-field theory  the limit $V \to \infty$
is carried out after the calculation of physical 
observable values:  scattering sections,
probabilities of decays and so on, which are
normalized on   time and volume units.}.\par
The vacuum density of  energy of the monopole solution:
\be
\label{dens}
\sim <B^2> \equiv \frac {1}{\alpha_s\epsilon V}=
\frac{4\pi m}{g\sqrt{\lambda} V},
\ee
is removed by the  finite counter-term
in the Lagrangian \cite {David2}:
\be
\label{contr}
{\bar{\cal L}}= {\cal L}-\frac {< B ^2 >}{2}.
\ee
This nonzero vacuum magnetic tension is a crucial difference of the topological
degeneration of  fields in  the Minkowski space from the topological
degeneration of the classical vacua in the  instanton theory  in the
Euclidean one(where the normalization of the vacuum, $F^{\mu \nu}=0$,
is a precondition of the formula (\ref {v}): see p.p.482-483 in \cite{Cheng}).
The Bogol'nyi equation (\ref {Bog}), applied for the vacuum topology
(\ref{top2}),  provides this 
nonzero vacuum magnetic tension.
\par
The problem is \it to formulate the Dirac quantization of  weak perturbations
of   non-Abelian  fields in the presence of the
non-perturbation monopole\rm, taking  account of  the topological
 degeneration of all the initial data.
\subsection{ Dirac method and Gribov copies.}
Instead of the artificial equations of the  gauge-fixation method \cite{Fadd3}:
\be
\label{fix}
F(A_\mu)=0,\quad F(A_\mu ^u)= M_F u \neq 0 \Longrightarrow Z^{FP}= 
 \int \prod \sb {\mu} D A_\mu det M_F \delta (F(A))e^{iW},
\ee
we repeat the Dirac \it constraint-shell \rm formulation resolving the
constraint (\ref{Gauss}) with  nonzero initial data:
\be
\label{init}
\partial _0 A_i ^c =0 \Longrightarrow  A_i ^c(t,{\bf x})=\Phi_i^{c(0)}({\bf x}).
\ee
The vacuum \it magneto-static \rm  field $\Phi_i^{c(0)}$ has its zero value
of the winding number (\ref{wind}),\linebreak $X[\Phi_i^{c(0)}]=0$,
and satisfies the classical equations everywhere besides of a
small region near the origin of coordinates of the size 
\be
\label{size}
\epsilon \sim \frac {1}{\int d^3x  B ^2 (\Phi)} \equiv \frac {1}{< B ^2 >V}
\ee
that disappears in the infinite volume limit.\par
The second step is the consideration of  the perturbation theory (\ref{sum}),
 where the constraint (\ref{Gauss}) acquires  the form
\be
\label{Gauss1}
[D^2 (\Phi ^{(0)})]^{ac}A_{0c}^{(0)}=\partial _0 [D^{ac}_i(\Phi ^{(0)})A_c^{i ~(0)}].
\ee
Dirac proposed \cite{Dir} to remove the time component $A_0$
(the quantization of which contradicts the quantum principles as  a
 \it non-dynamic
degree of freedom\rm); then the constraint (\ref{Gauss1}) acquires the form
\be
\label{Gauss2}
\partial _0 [D^{ac}_i(\Phi ^{(0)})A_c^{i~ (0)}]=0.
\ee
We define  the \it constraint-shell gauge \rm
\be
\label{cshg}
[D^{ac}_i(\Phi ^{(0)})A_c^{i ~(0)}]=0
\ee
as  zero initial data of this constraint. \par
It is easy to see that the expression in the square  brackets in
(\ref{cshg}) can be treated as \it an equal to zero, in the
initial  time instant, longitudinal component of a YM field \rm
(\ref {init}). We shall denote it as $A^{a\parallel}$:
\be
\label{Aparallel}
A^{a\parallel}\equiv [D^{ac}_i(\Phi ^{(0)})A_c^{i ~(0)}]=0\vert _{t=0}~.
\ee
Let us call the latter (\it initial  \rm ) condition as  the 
\it covariant Coulomb gauge\rm.
Then the constraint (\ref{Gauss1})
means that the  time derivatives of longitudinal
fields equal to zero. 
\par
The topological degeneration of initial data means \it that not only the
classical vacua, but also all the fields $A_i^{ (0)}=\Phi_i^{ (0)}+
{\bar A}_i^{ (0)}$ in the gauge \rm (\ref{cshg})  are degenerated\rm:
\be
\label{degeneration}
{\hat A}^{(n)}_i= v^{(n)}({\bf x}) ({\hat A}_i^{ (0)}+
\partial _i)v^{(n)}({\bf x})^{-1},\quad v^{(n)}({\bf x})=
\exp [n\Phi _0({\bf x})].
\ee
The winding number functional (\ref{wind}), after the transformation (\ref{cl.vac}),
takes the  form (see the formula (3.36) in\cite {David2})
\be
\label{chang}
X[A^{(n)}_i]= X[A^{(0)}_i]+{\cal N}(n)+\frac {1}{8\pi ^2}\int d^3x \epsilon^{ijk}
 Tr [\partial _i({\hat A}_j^{ (0)} L^n_k)],
\ee
where ${\cal N}(n)=n$ is given by the eq. (\ref{degree}).\par
The constraint-shell gauge (\ref{cshg}), (\ref{Aparallel}) keeps its form in each topological class:
\be
\label{transv}
D_i^{ab} (\Phi _k^{(n)}){\bar A}^{i(n)}_b =0 
\ee
if the  phase $\Phi _0({\bf x})$ satisfies
\it the equation of the Gribov ambiguity \rm  
for the constraint-shell gauge (\ref{cshg}), (\ref{Aparallel})
(see also \S T.26 in \cite {Al.S.}):
\be
\label{Gribov.eq}
[D^2 _i(\Phi _k^{(0)})]^{ab}\Phi_{(0)b} =0;
\ee
this leads to the zero FP determinant $det$ $(\hat \Delta)$ in (\ref{FP})
(it has the countable set of  eigenvalues $\lambda_i$ corresponding
to the zero solution (\ref{Gribov.eq}))). Note that \it the Gribov equation \rm(\ref{Gribov.eq}),
written down in terms of the Higgs isoscalar $\Phi_{(0)b}$,
is the direct consequence of the Bogomol'nyj equation in the form (\ref{Bog1}) and the
Bianchi identity $\epsilon ^{ijk}\nabla _i F_{jk}^b =0$.
\par
Note also that (although indirectly) the equations  
(\ref{Gauss2}), (\ref{Aparallel}) are  Cauchy  conditions for the 
Gribov ambiguity eq. (\ref{Gribov.eq}).\par
The Gribov ambiguity equation  has a very interesting geometric interpretation.
We should recall, to begin with, that (\S T.22 in \cite {Al.S.})
every gauge
field $A_\mu$, as an element of the adjoin representation of the given Lie
algebra, sets  some element $b_\gamma \in G$, where $G$ is the considered gauge
Lie group. These elements $b_\gamma$ are defined as
\be
\label{bg}
b_\gamma =P\exp (-\int {\sb \Gamma}  {\bf T} \cdot A _\mu dx^\mu),
\ee 
where $P$ is the symbol of the parallel transfer along the curve $\gamma$
in the coordinate (for example, Minkowski)space; $\bf T$ are the matrices of the
adjoin representation of the Lie algebra. It is obvious that
$b_\gamma =b_{\gamma_1} b_{\gamma_2}$ as the end of the curve
$\Gamma_1$ coincides with the beginning of the curve $\Gamma_2$
and the curve $\Gamma$ is formed from these curves. Thus, the group
operation (the multiplication) is associative; there exists always
the unit element and  element inverse to the given one.
This follows from the usual features of curves and the exponential 
function. \par
We see that elements $b_\gamma$ are defined on the set 
of  external 1-forms in the Lie algebra. The cohomology classes
of these external 1-forms are the elements of the cohomology group $H^1$
(see for example \S T.7 in \cite {Al.S.}).
The Pontryagin formula
for a degree of a map (see the lecture (26) in\cite{Postn3}):
\be
\label{deg}
\int \sb {X} f^*\omega = deg f
\ee
(where the map $f:X\to Y$ is smooth and maps 
the compact space $X$ into the compact space $Y$: the 
so-called \it eigen map\rm; $f^*$ is the 
homomorphism of the cohomology groups: $H^1_X\to H^1_Y$,
induced by the map $f$ ), sets an \it one-to-one correspondence
between the homotopy and cohomology groups  \rm in considered theory
\footnote{We should apply the formula (\ref {deg}) to the topology
(\ref{top2}) in our consideration.}.
In particular, the 
topological charge $n=0$ corresponds to \it exact \rm 1-forms, i.e to those
which can be represented as a differential, $d\sigma$, of some
1-form $\sigma$. Because of the \it Poincare lemma\rm,
$d\cdot d\sigma =0$, i.e. \it every exact form is closed\rm.
Note that the scalar field $\Phi_{(0)b}$ in the
Gribov equation (\ref {Gribov.eq}) has the topological charge $n=0$. This
charge ,through the \it smooth \rm  Bogomol'nyj equation, is told to the
magnetic tension tensor $F^b_{jk}$ (i.e to corresponding 2-forms) and to the 
corresponding  YM fields (1-forms).\par
Let us consider now those curves $\Gamma$ which begin and finish at
the same point $x_1$ of the Minkowski space. Such closed curves
are called the 1-\it dimensional cycles\rm, and the corresponding
elements $b_\gamma$ can be written down as  cyclic integrals: 
\be
\label{cycl}
b_\gamma =P\exp (-\oint {\sb\Sigma} {\bf T} \cdot A _\mu dx^\mu)~,
\ee
over some 1- dimensional cycle $\Sigma$. According to the
\it De Rham theorem \rm (see p.276 in \cite{Al.S.}), if
some external form $\omega$ is exact, its integral
over \it every \rm cycle defined on the considered manifold $M$
is equal to zero (this confirms also the formula (\ref{deg}) for the zero
topological class).
This means that the 
integral in (\ref{cycl}) is equal to zero ($b_\gamma =1$) \it for every exact form \rm 
(corresponding to the zero topological charge according (\ref{deg})).\par
The elements (\ref{cycl}) form \it the holonomy subgroup \rm $H$ in the 
initial gauge group. Those of these elements which
correspond to the exact 1-forms form, in  turn,  the \it
restricted holonomy subgroup\rm, which we  shall denote as
$\Phi^0$. The action $b_\gamma$ of the gauge group
$G$ on the manifold $M$ is described in terms of the principal fibre
bundle $P(M,G)$ over the
manifold $M$; thus, we shall write henceforth 
$\Phi^0(u)$ ($u \in P(M,G)$ is a fixed point of the contour $\Sigma$)
for the elements of  $\Phi^0$.  
\par
One can say that the unit element of the holonomy group
$H$ is \it degenerated \rm  with respect to all the exact
forms corresponding to the zero topological charge.
In our case of the vacuum Higgs fields $\Phi_{(0)b}$ namely
these fields determine the class of exact forms and the
group $\Phi^0(u)$.\par
As it was shown in the monographs \cite{Al.S.}, if 
$A_\mu$ and $ A'_\mu$ are some two gauge fields, the
gauge equivalence between which is realized by the 
function $g(x)\in G$, then
\be
\label{equiv}
b'_\gamma = g(x_1)b_\gamma g(x_1)^{-1}
\ee
as the curve $\Gamma$ begins and ends in the point $x_1$
(\it the holonomy elements of some two gauge equivalent fields are
conjugate\rm ).\par
Let $g(x)$ have the spatial asymptotic $g(x)\to 1$ as $x\to \infty$(indeed,
$g(x)$ would have such asymptotic already on distances $\sim 1$ fm.; this is connected
with needs of the quark confinement as we shall see this in Sections 8,9).
Let also $b_\gamma$ and $b'_\gamma$ be  elements of $H$
constructed by external forms belonging to some one class of
cohomologies. In conclusion, the gauge fields $A_\mu$ and $ A'_\mu$, 
forming the elements $b_\gamma$ and $b'_\gamma$ ,respectively, and
connected by the gauge transformation $g(x)\to 1$,
\it have the Coulomb  gauge \rm (\ref{Aparallel}). Every such class
is obtained from the zero class
of exact forms \it as its Gribov copy \rm (therefore as a 
Gribov copy of the ambiguity equation(\ref{Gribov.eq})also).  
As 
$g(x)\to 1$, $x\to \infty$, we can rewrite (\ref{equiv}) as
\be
\label{coh}
b'_\gamma =b_\gamma \cdot 1 ~.
\ee
The latter equality reflects the structure of the
cohomology group $H^1$: \it some two \rm 1-\it forms
belonging to one class of
cohomologies are equivalent to within some
exact form \rm (\S T6 in \cite {Al.S.}). In terms
of the holonomy group $H$ this means that
some two elements of $H$ corresponding to the 1-forms
belonging to one class of
cohomologies are \it equivalent  to within some 
element of the restricted holonomy group $\Phi^0$ \rm. \par
Thus, the Gribov equation (\ref{Gribov.eq}) and the Gribov
transformations (\ref{degeneration})
describe correct the 
cohomologic structure of the transverse vacuum YM fields (satisfied the 
Coulomb  gauge \rm (\ref{Aparallel})) as elements of
the connection of the principal fibre bundle $P(M,G)$, with the 
structural non-Abelian group $SU(2)$  broken down spontaneous to
the $U(1)$ group and the Minkowski space $M$ as a base of 
this fibre bundle, at the spatial infinity.\par
Let us  prove \it that the holonomy  group \rm $H$ \it has a nontrivial
structure  in the non-Abelian case only \rm (\S T.26 in \cite {Al.S.}). \par
Let us consider   the \it principal trivial fibre bundle
\rm $\zeta =(M\times G,M,G,p)$ of  \it topological trivial gauge fields
on some manifold \rm $M$; for example, it is the Minkowski space as in our
theory (a more detailed discussion of this class of  fibre bundles
is given in the monograph \cite{Postn4}).  The \it total \rm  space ${\cal E}_0=M\times G$
of  this fibre bundle can be
deformed at a point. It follows from the fact that  trivial fibre
bundles are described (see the example 1 on p.107 in \cite {Postn4})
with the help of  the unique  chart map ( the \it trivialisation\rm):
\be
\label{triv}
M\times R^n\to M\times  \cal V ,
\ee
where $ \cal V$ is some linear space over some field $K$.\par
In this case\cite {Switz} all the homotopy groups are trivial: $\pi _n {\cal E}_0 =0$;
this means that $ {\cal E}_0 =0$ can be deformed at a point. \par
The nontrivial topology appears after the identification of all the
 gauge equivalent fields. Let us mark out the subgroup
$G_0^\infty \subset G^\infty$ consisting of  transformations
defined with  functions $g(x)$ such that $g(x_0)=1$($x_0 \in M$).
This group acts \it free \rm on $ {\cal E}_0 $.
Really,
the transformation law of the element  $b_\gamma$ under the gauge
transformation corresponding to $g(x)$ has the form
\be
\label{b}
b'_\gamma= g(x_2)b_\gamma g(x_1)^{-1},
\ee
(compare with (\ref{equiv})),
where $x_1$ and $x_2$ are the beginning and  end of  the curve $\gamma$.
This law means that the group $G_0^\infty $ acts  free \rm on
$ {\cal E}_0 $. Really, if a gauge transformation keeps a gauge field
$A$ on its place, then
\be
\label{b1}
b_\gamma= g(x_2)b_\gamma g(x_1)^{-1}.
\ee
Choosing the curve to begin at $x_0$ and to end at an arbitrary point $x$,
we see that $b_\gamma= g(x)b_\gamma$, i.e. $g(x)\equiv 1$. This is the
definition of a free action of a group on a manifold
(see p.16 in \cite {Postn4}).\par
But the group $G^\infty$ acts not free on ${\cal E}_0$ in the general case,
and this is, as we shall  show now, a ground cause of the Gribov
ambiguities in the
non-Abelian case. For example, if some gauge field $A_\mu$ takes its values
in the Lie algebra ${\cal H}'$ of the  group $ H' \subset G$, and
$g\in G$ is an element commuting with all the elements of $H'$,
then such gauge transformations are  induced with $g(x)\equiv g$ (the  global
gauge transformations).\par
The holonomy  group $H$  is an example of such groups: there exists always
such gauge transformation $g(x)$ that maps an element 
$b_ \gamma \in H$ to itself.\par
The space ${\cal B}_0$ of  orbits of $G^\infty_0$ in ${\cal E}_0$ is, obviously,
the base of  the fibre bundle with the  fibre $G^\infty_0$ and
total space ${\cal E}_0$. 
The task of the choice of the \it unique \rm  gauge field
from every orbit (\it the gauge fixation \rm in the language of the  FP theory
of  path integrals) is equivalent to the task of the
construction of a \it  section \rm of the above  fibre bundle.
In the case if the fibre bundle has a section every element of the
homotopy
group of the base is obtained by a natural homomorphism from
some element of the homotopy group of the total space ${\cal E}_0$
(more precisely,  some element $\alpha$ is obtained from the element
$q\alpha$, where $q$ is the section). Since, as we  already emphasized,
$\pi_i {\cal E}_0=0$ at every $i$, $\pi_i {\cal B}_0=0$ at every $i$ (this follows
 from the definition of  homomorphism between the
Hausdorff  spaces ${\cal E}_0$ and ${\cal B}_0$). \par
On the other hand (see p.322 in \cite {Postn4}), if
\be
\label{hom.ch}
\pi_k {\cal E}_0=\pi_{k-1} {\cal E}_0=0,
\ee
then
\be
\label{pi.bas}
\pi_k {\cal B}_0= \pi_{k-1}G^\infty_0.
\ee
This is the result of the statement that (p.454 in \cite {Postn4}) the homotopic
 sequence of some fibre bundle with the linear-connected
total space $\cal E$ and the base $\cal B$:
\be
\label{pi1}
\pi_1 {\cal }E =\pi_1{\cal B} =0,
\ee
having the form
\be
\label{sec}
... \to \pi_{n+1}{\cal B}\longrightarrow  \sb{\partial} \pi_n {\cal F}
\longrightarrow \sb{ i_* }\pi_n {\cal E}\longrightarrow \sb { p_*} \pi_n {\cal B} \to ...
\ee
(where $\cal F$ is the fibre of the considered fibre bundle), is \it exact\rm. \par
It follows from (\ref{pi.bas}) that the section exists  if and only if
all  the homotopy
groups of  the space $G^\infty_0$ are trivial. If the manifold $M$ is 
topological equivalent  to the sphere $S^n$, the homotopy  group $\pi_k G^\infty_0$ is
isomorphic to the group $\pi_{k+n}G$. Really, in this case we can
identify $G^\infty_0$ with the space of maps of the cube $I^n$ in $G$
transferred all  the bound $\dot I^n$ of  the cube $I^n$ in  the unit element of the  group $G$
(it follows from the remark that the sphere $S^n$ is obtained from $I^n$ by
 the deformation of the whole bound at a point ). The
elements of  the homotopy group $\pi_{k}G^\infty_0$ can be represented as
 homotopic classes of   maps of the cube $I^k$ in
$G^\infty_0$ transferred its bound $\dot I^k$ in the unit element of the
group $G^\infty_0$. This map associates a function $g_\nu(x)$
with the
values in $G$ defined on $I^n$ and satisfying the  condition
$g_\nu=1$ if $\nu\in\dot I^k$; $x\in I^n$, or $\nu\in I^k; x\in\dot I^n$.
Considering
the
pair $(x,\nu)\in I^n \times I^k$ as a point of  the cube $I^{n+k}$, we
see that the maps $I^k\to G^\infty_0$ is in an one-to-one correspondence
with the maps $I^{n+k} \to G$ transferring all its bound,
\be
\label{bound}
\dot I^{n+k}= \dot I^n \times I^k \bigcup I^n \times \dot I^k~,
\ee
in the unit element of  the group $G$. The homotopic classes of  maps of
$I^{n+k}\to G $, by definition, form the group $\pi_{n+k}G$.
This means that
$\pi_{k}G^\infty_0$ is \it isomorphic \rm to $\pi_{n+k}G$.\par
If $G$ is a \it compact non-Abelian group\rm, we can prove that it is
impossible to choose, in a continuous way, the one gauge field from
every
orbit of $ G^\infty_0$ on $S^n,n>0$, i.e.
\it it is impossible to fix a gauge removing completely the gauge freedom\rm. \par
In the case of $S^3$ and if $G=SU(2)$ this follows immediately from the relation
\be
\label{pi0}
\pi_0 G^\infty_0 =\pi_3 G=\pi_3 SU(2)=\bf Z
\ee
(see (\ref{3-path})). The general proof (for an arbitrary compact non-Abelian group)
is to check that a compact non-Abelian group $G$ has
nonzero homotopy subgroups in arbitrary dimensions.\par
The proved results are connected with the problem of the continuous gauge
fixation on the orbits of $ G^\infty_0$. However, it is easy
to study the question about the continuous gauge fixation on the orbits
of $ G^\infty$, utilising the same methods. The groups $ G^\infty$
act not free on ${\cal E}_0$ in the general case. But, removing the fields
with  the holonomy group $H$ which do not coincide with $G$ we obtain
the
subspace of  fields of a common position ${\cal E}'_0$, on which
$ G^\infty$ \it acts free\rm. The removal of the submanifold given
by the infinite number of the equations does not influence  homotopic
groups, therefore
\be
\label{E'}
\pi_i {\cal E}'_0=0,\quad i\geq 0.
\ee
This allows us to ascertain \it the absence of any section \rm of the fibre
bundle with  the total space ${\cal E}'_0$ in the case when gauge
fields are defined on $S^n,n>0$, and take theirs values in the Lie algebra
of the compact non-Abelian group $G$.\par
The gauge
fields on  the Minkowski space, by the condition of  the spatial asymptotic of the
(\ref{asym}) type, can be considered as those defined on $S^n,n>0$. \it
This is the cause of the Gribov ambiguities in the non-Abelian case\rm.\par
We considered above the case of  topological trivial gauge fields
defining the connection of the trivial principal fibre bundle $\zeta$.
All these
results can be extended to the case of  topological nontrivial gauge
fields (i.e.\it to connections of  principal fibre bundles\rm).
For example,  one can consider, instead of ${\cal E}_0$, the space ${\cal E}_n$
of gauge fields with the topological number $n$ on the  sphere $S^4$.
The space ${\cal E}_n$  can also  be deformed at a point (this deformation is
defined as $A_t=(1-t)A+t A^{(0)}$).\par
Such is the origin
of the Gribov ambiguities.\par
One can show \cite {David1,David2} that  the Gribov equation (\ref{Gribov.eq}), together with
the topological condition
\be
\label{X[n]}
X[\Phi ^{(n)}]=n,
\ee
are compatible with the unique solution of the classical equations.
\it It is just the Wu-Yang monopole considered before\rm.
The nontrivial solution of the equation for the Gribov phase (\ref{Gribov.eq})
in this case is well-known:
\be
\label{phase}
{\hat \Phi}_0= -i\pi \frac {\tau ^a x_a}{r}f_0^{BPS}(r),
\quad f_0^{BPS}(r)=[\frac{1}{\tanh (r/\epsilon)}-\frac{\epsilon}{r} ].
\ee
It is the Bogomol'nyi-Prasad-Sommerfeld (BPS) monopole \cite {BPS}.
 Note here the very important detail. The Gribov phase (\ref{phase})
is nothing else than the $SU(2)$ \it   Higgs isoscalar \rm (compare 
with the formula (23) in the paper \cite{David1}).\par
%To make our reader certain  of this fact, we refer him  to the monograph
%\cite {4} (\S 11, the formula (10)) and  the paper\cite {33}
%the formula (22),(23)).
%This means that \it the  topological degeneration
%of  classical vacuum is realised due to the Higgs sector of  the theory
%and depends explicitly on the vacuum expectation value $a$ of  Higgs scalar
%field.\rm \par
Thus, instead of  the topological degenerated classical vacuum for the
instanton theory (in the physically unattainable region),
we have \it  the topological degenerated Wu-Yang monopole\rm:
\be
\label{mon.deg}
{\hat \Phi_i} ^{(n)}:= v^{(n)}({\bf x})[{\hat \Phi_i} ^{(0)}+\partial _i]v^{(n)}({\bf x})^{-1},\quad
v^{(n)}({\bf x})=
\exp [n\hat \Phi _0({\bf x})],
\ee
and the topological degenerated multipoles:
\be
\label{mult}
{\hat {\bar A}}^{(n)}:=v^{(n)}({\bf x}){\hat {\bar A}}^{(0)}v^{(n)}({\bf x})^{-1}.
\ee
The Gribov copies are an evidence of the zero mode in the left-hand side of
 both the constraints (\ref{Gauss}),(\ref{Gauss1}):
\be
\label{zero mode}
[D^2_i(\Phi ^{(0)})]^{ac} A_{0c}=0.
\ee
The nontrivial solution of this equation,
\be
\label{sol.zero}
A_0^c(t,{\bf x})= {\dot N}(t)\Phi_0^c ({\bf x}),
\ee
can be removed from the local equations of motion by the gauge transformation (a la Dirac of 1927) to convert
the  fields \it into the
Dirac variables\rm:
\be
\label{d.v.}
\hat A_i^{(N)}=\exp [N(t)\hat \Phi_0 ({\bf x})]
[\hat A^{(0)}_i+\partial _i]\exp [-N(t)\hat \Phi_0 ({\bf x})].
\ee
But the solution (\ref{sol.zero}) cannot be removed from the constraint-shell action 
$W^*= \int dt {\dot N}^2 I/2+...$ and from the winding number
$X[A^{(N)}]=N+X[A^{(0)}]$. Finally, we obtain the Feynman path integral
\be
\label{path i}
Z_F=\int DN \prod \sb {i,c}  [D E^{c(0)}_{i} D A^{i(0)}_{c}]e^{iW^*}
\ee
that contains the additional topological variable.\par
We consider the derivation of the integral (\ref{path i}) in the next sections.
\subsection{ Topological dynamics and chromo-electric monopole.}
The repetition of the Dirac definition of observable variables in QED
allowed us to determine  vacuum  fields and  phases of their
topological degeneration in  the form of  Gribov copies of  the  constraint-shell
 gauge.\par
The degeneration of  initial data is an evidence of  the  zero mode of the
 Gauss law constraint. In the lowest order of the considered
perturbation theory the constraint (\ref{zero mode}) has the solution (\ref{sol.zero})
with the  \it electric  monopole \rm 
\be
\label{el.m}
F^b_{i0}={\dot N}(t)D ^{bc}_i(\Phi_k ^{(0)})\Phi_{0c}({\bf x}).
\ee
We call this new variable $N(t)$ - the \it winding number variable\rm,
 and it is defined by the \it vacuum Chern-Simons functional\rm,
which is equal to the difference of the \it in \rm and  \it out \rm
values of this variable:
\bea
\label{winding num.}
\nu[A_0,\Phi^{(0)}]&=&\frac{g^2}{16\pi^2}\int\limits_{t_{in} }^{t_{out} }dt
 \int d^3x F^a_{\mu\nu} \widetilde{F}^{a\mu \nu}=\frac{\alpha_s}{2\pi}
 \int d^3x F^b_{i0}B_i^b(\Phi^{(0)})[N(t_{out}) -N(t_{in})]\nonumber \\
 &&
 =N(t_{out}) -N(t_{in})~.
\eea 
The winding number functional admits its generalization \it to the
noninteger degrees of the map \rm \cite {Niemi}:
\be
\label{Xni}
X[\Phi^{(N)}]\neq n (n \in {\bf Z}),\quad  ({\hat \Phi}^{(N)}=
e^{N{\hat \Phi}_0}[\hat \Phi_i ^{(0)}+\partial _i]e^{-N{\hat \Phi}_0}).
\ee
Thus, we can identify the global variable $N(t)$ with the winding number of
degrees of freedom in the Minkowski space described by the \it free rotator action \rm
\be
\label{rot}
W_N=\int d^4x \frac {1}{2}(F_{0i}^c)^2 =\int dt\frac {{\dot N}^2 I}{2},
\ee
where the momentum of rotator (see also (4.6) in  \cite{David2}):
\be
\label{I}
I=\int \sb {V} d^3x (D_i^{ac}(\Phi_k)\Phi_{0c})^2 =
\frac {4\pi^2\epsilon}{\alpha _s}
=\frac {4\pi^2}{\alpha _s^2}\frac {1}{ V<B^2>}~,
\ee
 does not
contribute  to the local equations of motion. The
free rotator action disappears in the limit $V\to \infty$. \par
Now, with account of the evaluation of the magnetic energy,
(\ref{magn.e}), we can write down the action of the YM
theory in the Minkowski space in the lowest order. This  
action contains as an "electric" as a "magnetic" BPS 
monopoles:
\be
\label{ca}
W_{\cal Z} [N,\Phi^{0 BPS}]= \int dt d^3x \frac {1}{2}\{ [F^b_{i0}]^2-
[B^b_i(\Phi^{0BPS})]^2\}= \int dt\frac {1}{2}\{ I\dot N ^2- 
\frac {4\pi}{g^2 \epsilon}\}~.
\ee 
The topological degeneration
of all the
fields reduces to the degeneration of only one global topological
variable $N(t)$ with respect to the shift of this variable on integers:
$N \Longrightarrow N+n,n= \pm 1,\pm 2,...;0\le N(t)\le 1$.
Thus, the topological variable $N(t)$
determines the  free rotator \rm with the instanton-type wave function (\ref{plan})
of the topological motion in
the Minkowski space-time:
\be
\label{PsiN}
\Psi _N=\exp (iP_N N), \quad P_N ={\dot N} I= 2\pi k +\theta ,
\ee 
where $k$ is  the \it number of the Brillouin zone\rm,
and $\theta$ is the $\theta$-angle (or the \it  Bloch quasi-momentum\rm)
\cite{Pervush1}.The action (\ref{ca})of the YM
theory in the Minkowski space in the lowest order induces
the corresponding Hamiltonian (in terms of the canonical 
momentum $ P_N =\dot N I$):
\be
\label{Hamilton}
H= \frac {2\pi}{g^2\epsilon}[ P_N^2 (\frac {g^2}{8\pi^2})^2+1].
\ee
 In contrast to the instanton wave
function (\ref{plan}),
the spectrum of the topological momentum is \it real \rm and belongs to
the \it physical values\rm. Finally, the equations (\ref{rot}),(\ref{PsiN})
determine \it the  countable spectrum \rm of the global electric tension (\ref{el.m}):
\be
\label{el.m1}
F^b_{i0} ={\dot N}[D_i(\Phi ^{(0)})A_0]^b = \alpha _s (\frac {\theta}{2\pi}+k)B^b_i(\Phi ^{(0)}).
\ee
It is an analogue of the Coleman spectrum of the electric tension in the
$QED_{(1+1)}$ \cite {Coleman}:
\be
\label{Col.sp}
G_{10}=\dot N \frac {2\pi}{e} =e(\frac {\theta}{2\pi}+k).
\ee
The application of the Dirac quantization to the 1-dimensional
electrodynamics $QED_{(1+1)}$ in the paper \cite{Gogilidze}
demonstrates the universality of the Dirac variables and their
adequacy to the description of the topological dynamics in terms of
a nontrivial homotopy group.
\section {Zero mode of Gauss Law.}
\subsection {Dirac variables and zero mode of Gauss Law.}
The constraint-shell theory is obtained by the explicit resolution of the
Gauss law constraint (\ref{Gauss}), and our  next step is connected with the
initial action on the surface of these solutions:
\be
\label{in.act}
W^*= W[A\mu] \vert _ {\frac {\delta W}{\delta A_0}=0} .
\ee
The results of a similar solution in QED are \it  electrostatic and
 Coulomb-like atoms\rm. In the non-Abelian case the
topological degeneration in the  form of Gribov copies means that
the  general solution of the Gauss law constraint  (\ref{Gauss}) contains the zero
mode ${\cal Z}$. The general solution of  the  inhomogeneous equation (\ref{Gauss}) is
the sum of the zero-mode solution ${\cal Z}^a$
of the homogeneous equation,
\be
\label{hom}
(D^2 (A))^{ab}{\cal Z}_b =0 ~,
\ee
and a particular solution, ${\tilde A}_0^a$, of the inhomogeneous one:
\be
\label{inhom}
A_0^a={\cal Z}^a+{\tilde A}_0^a.
\ee
 The zero-mode ${\cal Z}^a$, at  the spatial infinity, can be represented
 as a form of the sum of the product of the new topological variable
${\dot N}(t)$,  Gribov phase $\Phi _{(0)}({\bf x})$ and weak
multipole corrections:
\be
\label{Summ}
{\hat {\cal Z}}(t,{\bf x})\vert _{asymptotic}= {\dot N}(t)\hat\Phi _{(0)}({\bf x})+O (\frac {1}{r^{l+1}}),
\quad (l>1).
\ee
In this case the single one-parametric variable $N(t)$ reproduces the
topological degeneration of all the  field variables,  if the Dirac variables
are defined by the gauge transformations
\be
\label{tr.d}
0= U_{\cal Z}({\hat {\cal Z}}+ \partial _0)U_{\cal Z}^{-1}
\ee
(\cite {David2}, the formula (2.9)),
\be
\label{tr.d1}
{\hat A}^*_i =U_{\cal Z}({\hat A}_i^0+\partial _i)U_{\cal Z}^{-1}, \quad 
{\hat A}_i^{(0)} =\Phi _i^{(0)}+{\bar A}_i^{(0)},
\ee
where the spatial asymptotic of $U_{\cal Z}$ is ((2.14) in \cite {David2})
\be
\label{spat.as}
U_{\cal Z}= T \exp [ \int ^t dt'{\hat {\cal Z}}(t',{\bf x})]\vert _{asymptotic}=
\exp [N(t)\hat\Phi _{(0)}({\bf x})].
\ee
The topological degeneration of all  the fields reduces to the
degeneration of only one global topological variable $N(t)$ with
respect to the
shift of this variable on integers:\linebreak $(N \Longrightarrow N+n,n= \pm 1,\pm 2,...)$.
\subsection { Constraining with  zero mode.}
Let us formulate the equivalent unconstrained system for the YM theory in
the monopole class of functions in the presence of the zero mode
${\cal Z}^b$ of the Gauss law constraint:
\be
\label{A}
A_0^a={\cal Z}^a+{\tilde A}_0^a ;\quad F^a_{0k}=
- D^{ab}_k(A){\cal Z}_b+{\tilde F}^a_{0k} \quad (( D^2(A)^{ab}{\cal Z}_b =0 ).
\ee
To obtain the constraint-shell action:
\be
\label{constr}
W_{YM}(constraint) = {\cal W}_{YM}[{\cal Z}]+{\tilde W}_{YM}[{\tilde F}],
\ee
we use the obvious decomposition:
\be
\label{relations}
F^2=(-D{\cal Z}+{\tilde F})^2= (D{\cal Z})^2- 2{\tilde F}D{\cal Z}+{\tilde F}^2= 
\partial ({\cal Z}\cdot D{\cal Z})-
2 \partial ({\cal Z}{\tilde F})+({\tilde F})^2.
\ee
The latter relation is true because of the Bianchi identity $D {\tilde F}=0$,
the Gribov equation $D ^2{\cal Z}=0$ and the explicit expression for the
derivative $D{\cal Z}$:
$D{\cal Z}= (\partial {\cal Z}+ g {\bf A}\times{\bf {\cal Z}})$.
This shows that the zero mode part, ${\cal W}_{YM}[{\cal Z}]$, of
the constraint-shell action (\ref{constr}) is the sum of the two integrals:
\be
\label{int.sum}
{\cal W}_{YM}[{\cal Z}]= \int dt \int d^3x [\frac {1}{2} \partial_i ({\cal Z}^a D^i_{ab}(A){\cal Z}^b)-
\partial_i (\tilde F^a_{0i}
{\cal Z}_a)] ={\cal W}^0+{\cal W} ',
\ee
where  the first term,${\cal W}~^0$, is  the \it  action of a free rotator,
\rm and the
second one, ${\cal W} ~'$,
describes \it the coupling of the zero-mode to  local excitations\rm.
These  terms are determined by the asymptotic of  fields
$({\cal Z}^a,A^a_i)$ at the spatial infinity: (\ref{Summ}), (\ref{asym}).
We denote them as ${\dot N}(t)\Phi _{(0)}^a({\bf x}),\Phi _i^a({\bf x})$.
The fluctuations $\tilde F^a_{0i}$ belong  to the class of multipoles.
Since the  integral over  monopole-multipole couplings vanishes
(the Gauss-Ostrogradsky theorem and the asymptotic (\ref{Summ})), the fluctuation part
 of the second term  drops out. The substitution of
the solution with
the asymptotic (\ref{Summ})) into  the first  term of the eq. (\ref{int.sum}) leads to the
 zero-mode action (\ref{rot}).\par
The action for the equivalent unconstrained system of local excitations
 (compare with the formula (21) in \cite {Pervush2}):
\be
\label{unconstr}
\tilde W_{YM}[\tilde F]= \int d^4x\{ E_k^a \dot A_a^{k(0)}- 
\frac {1}{2} \{E_k^2+B_k^2 (A^{(0)})+[D_k^{ab}(\Phi^{(0)})\tilde \sigma_b]^2\}\}~,
\ee
is obtained in terms of  variables with the zero degree of the  map:
\be
\label{zero}
\hat{\tilde F}_{0 k}=U_{\cal Z}{\tilde F}_{0 k}^{(0)}U_{\cal Z}^{-1},\quad 
\hat{A}_i=U_{\cal Z}(\hat{A}_i^{(0)}+\partial_i)U_{\cal Z}^{-1},
\quad  \hat{A}_i^{(0)}(t,{\bf x})=\Phi_i^{(0)}(t,{\bf x})+\hat{\tilde A}^{(0)}(t,{\bf x})~,
\ee
by the decomposing  of the electrical components of the  field strength
tensor $F _{0 i}^{(0)}$ into their  transverse: $E_i^a$, and longitudinal:
$F _{0 i}^{aL}=- D_i^{ab}(\Phi ^{(0)})\tilde \sigma_b$, parts, so that
\be
\label{F0}
F _{0 i}^{a(0)}=E_i^a-D_i^{ab}(\Phi ^{(0)})\tilde \sigma_b.
\ee
Here the function $\tilde \sigma^b$ is determined from the Gauss equation
\be
\label{G.e}
((D^2(\Phi ^{(0)}))^{ab}+ g\epsilon ^{adc}\tilde A _{id}^{(0)}D^{ib}_c(\Phi ^{(0)}))\tilde \sigma_b =- g\epsilon ^{abc}
\tilde A _{ib}^{(0)}E^i_c
\ee
(we can recommend our reader the monograph \cite{Slavnov}, p.88,
where the formulas (\ref{F0}),(\ref{G.e}) were derived in the Hamiltonian formalism of the YM
theory). \par
If we introduce the current $j$ of independent non-Abelian
variables:
\be
\label{cur1}
j_0 ^a = g \epsilon ^{abc} [A_{ib}-\Phi^{a~(0)}]E ^{i}_c~,
\ee
the eq. (\ref{G.e}) can be rewritten as
\be
\label{cur-t}
D_i^{cd}(A)D^i_{db}(\Phi^{(0)})\tilde \sigma ^b = j_0 ^c~.
\ee
The latter equation depends in fact on the zero mode $\hat{\cal  Z}$ described
in the previous subsection. \par
Due to the gauge-invariance, the dependence of the action for local
excitations on the zero mode
disappears, and we get the ordinary generalization of the
 covariant  Coulomb gauge \cite {Schwinger,Fadd1,Gitman}
 in the presence of the  Wu-Yang monopole.
\section { Rising potential induced by monopole.}
Now we can calculate the Green function of the Gauss equation (\ref{Gribov.eq})
(see\cite {David2},\S 4.C):
\be
\label{Gr.eq}
D^2((\Phi ^{(0)})^{ab}({\bf x})G_b^c ({\bf x},{\bf y})=
\delta^{ac}\delta^3(x-y)
\ee
(it is the the Green function of the equation (\ref{cur-t}) simultaneously),
that forms the  potential of the current-current instantaneous interaction:
\be
\label{cint}
-\frac {1}{2} \int \sb{V_0} d^3 x d^3 y j~_0 ^b ({\bf x})
G_{bc} ({\bf x},{\bf y})j~_0^c({\bf y})~.
\ee
In the presence of the Wu-Yang monopole we have 
\be
\label{Gr.eq.mon}
D^2((\Phi ^{(0)})^{ab}({\bf x})
= \delta^{ab}\Delta -\frac {n^a n^b+\delta^{ab}}{r^2}+2(\frac {n^a}{r}\partial ^b-\frac {n^b}{r}\partial ^a),
\ee
where $n_a(x)=x_a/r$; $r=\vert {\bf x}\vert$.
Let us decompose $G^{ab}$  into the complete set of  orthogonal
vectors in the colour space:
\be
\label{complete set}
G^{ab}({\bf x},{\bf y})= [n^a(x) n^b(y)V_0(z)+ \sum \sb {\alpha=1,2} 
e^a_ \alpha (x)e^{b\alpha}(y)V_1(z)];\quad (z=\vert {\bf x}-{\bf y }\vert).
\ee
Substituting the latter into the first equation, we get
\it the Euler equation \rm (see\cite {Kamke}, the equation (2.160)):
\be
\label{Euler}
\frac {d^2}{dz} V_n+ \frac {2}{z}\frac {d}{dz}V_n- \frac {n}{z^2}V_n =0, \quad n=0,1 .
\ee
The general solution for the latter equation is
\be
\label{V}
V_n (\vert {\bf x}-{\bf y} \vert)=d_n\vert {\bf x}-{\bf y} \vert ^{l^n_1}+c_n\vert {\bf x}
-{\bf y} \vert^{l^n_2}, \quad n=0,1 ,
\ee
where $d_n,~c_n$ are constants, and $l^n_1,~l^n_2$
can be found as roots of the equation $l^{n2}+l^n=n$, i.e.
\be
\label{roots}
l^n_1= -\frac {1+\sqrt{1+4n}}{2};~~~~~\quad l^n_2=\frac {-1+\sqrt{1+4n}}{2}.
\ee
It is easy to see that for $n = 0$ at $d_0=-1/4\pi$ we get
 the \it Coulomb-type potential\rm:
\be
\label{Coulomb}
l^0_1= -\frac {1+\sqrt{1}}{2}=-1 ;\quad l^0_2=\frac {-1+\sqrt{1}}{2}=0,
\ee
\be
\label{Cp}
V_0 (\vert {\bf x}- {\bf y} \vert) = 
-1/ 4\pi \vert {\bf x}- {\bf y} \vert ^{-1} + c_0~;
\ee
and for $n = 1$, the "\it golden section \rm " potential with
\be
\label{ris}
l^1_1= -\frac {1+\sqrt{5}}{2}\approx -1.618;\quad l^1_2=\frac {-1+\sqrt{5}}{2}\approx 0.618 ,
\ee
\be
\label{ris1}
V_1 (\vert {\bf x}-{\bf y} \vert)=
-d_1\vert {\bf x}-{\bf y} \vert ^{-1.618}+c_1\vert {\bf x}-{\bf y} \vert^{0.618}~.
\ee
The latter potential (in  contrast with the Coulomb-type one)
can lead to the rearrangement of the naive perturbation series and to the
spontaneous chiral symmetry break-down. This potential can be considered
 as the origin of  "\it hadronization\rm " of  quarks and
gluons in QCD \cite {David2,Werner}.
\section {Feynman and FP path integrals.}
The Feynman path integral over  independent variables includes the  integration
over the topological variable $N(t)$:
\be
\label{N.i}
Z_F[J]= \int \prod \sb {t} dN(t)\tilde Z[J^U],
\ee
where
\be
\label{Z}
\tilde Z[J^U]=\int \prod \sb {t,x} \{\prod \sb {a=1} ^3 \frac {[d^2 A_a^{(0)}d^2 E_a^{(0)}]}{2\pi} \}
\exp i\{{\cal W}_{YM}({\cal Z})+\tilde W_{YM}(A_a^{(0)})+S[J^U]\}.
\ee
As we have seen above, functionals $\tilde W,~S$
are given in terms of  variables containing  non-perturbation
phase factors $U=U_{\cal Z}$, (\ref{spat.as}),
of the topological degeneration of  initial data.
These factors disappear in the action $\tilde W$, \it but not in
the source term\rm:
\be
\label{s.t}
S[J^U]=\int d^4x J^a_i \bar A^i_a,\quad \bar {\hat A}_i=
U(\hat A ^{(0)})U^{-1}~,
\ee
which reflects the fact of the topological degeneration of
physical fields. In general, the phase factors $U_{\cal Z}$,
as a relic of the fundamental quantization, remember all the 
information about the frame of reference, monopoles, 
rising potential of the instant interaction and other
initial data, including their topological degeneration 
and confinement (see farther).
\par
The constraint-shell formulation distinguishes the
\it bare\rm  "gluon", as a \it weak deviation of the monopole
with the index \rm $n = 0$,
and the
\it observable \rm (\it physical\rm ) "gluon"
\it averaged over the topological degeneration \rm (i.e.,
\it the  Gribov copies\rm) \cite {Pervush3}:
\be
\label{aver}
\bar A^{phys}= \lim \sb {L\to \infty} \frac {1}{2L} \sum \sb {n=-L} ^{n=+L}\bar A^{(n)}({\bf x})
\sim \delta_{r,0};
\ee
whereas in QED the constraint-shell field is
\it a transverse photon\rm.
A more detailed analysis of the latter formula will be conducted in
Section 8.\par
We can say that the Dirac variables with the topological degeneration
of  initial states in the  non-Abelian theory determine the
physical origin of hadronization and confinement as
non-local monopole effects. The Dirac variables distinguish the
unique gauge. In QED
\it it is the Coulomb gauge\rm;
whereas  in the YM theory \it it is 
the covariant generalization of the covariant  Coulomb gauge in
the presence of the monopole\rm.\par
If we pass to other gauges of  physical sources at the level of the
FP integral in  relativistic gauges, all the monopole effects
of the degeneration and rising potential can be lost
(as the Coulomb potential is lost in QED in  relativistic invariant
gauges). Recall that to prove the equivalence of the Feynman integral to
the Faddeev-Popov integral in an arbitrary gauge, we (\cite {Pervush2}, \S.2.5)
change variables and
concentrate all the monopole effects in  phase factors
(\cite {Pervush2}, the formula (40)) before  physical sources.
The change of  sources removes all these
effects (see\cite {Pervush2}, \S.2.3).\par
The change of  sources was possible in the Abelian theory only for
scattering amplitudes \cite {Fadd1} in  neighbourhoods
of poles of their Green functions 
when all the particle-like excitations of fields are on theirs mass-shells
(we recommend our reader to understand this fact with the example of
electron propagators).
However, for the cases of non-local bound states and other phenomena
where these  fields are \it off  their mass-shell\rm, the Faddeev
theorem about the
equivalence of the different "gauges"
(see for example (7.23) in \cite {Gitman}) is not valid.
\section {Free rotator: topological confinement.}
The topology can be an origin of  the colour confinement as
the complete destructive interference of  the phase factors of
the topological
degeneration of initial data.\par
The mechanical analogy of the topological degeneration of  initial
data is the free rotator $N(t)$ with  the action of a free particle
(compare with (\ref{rot}))
\be
\label{rot1}
W(N_{out},N_{in}\vert t_1)= \int \sb {0}  ^{t_1}dt \frac {\dot N ^2}{2}I,\quad p=
\dot N I,\quad H_0=\frac {p^2}{2I}
\ee
given on the ring, where the points $N(t) + n$ ($n \in \bf Z$)
are physically equivalent (see (\ref{PsiN})). Instead of  initial data
$N(t = 0) =N_{in}$
in  mechanics in the space with a trivial topology,
the observer of the rotator has a \it  manifold of  initial data \rm
$N^{(n)}(t=0)=N_{in}+n;~n=0,\pm 1,\pm 2,...$\par
The observer does not know  where is the rotator.
It can be at  points $N_{in}$, $N_{in}\pm 1,N_{in}\pm 2,...$.
Therefore,
he should  \it average \rm  the wave function (compare with (\ref{aver})):
\be
\label{Psi2}
\Psi (N)=e^{ipN}~,
\ee
over all the values of the topological degeneration with the
$\theta$-angle measure: $\exp (in\theta)$.
As a result, we obtain the wave function
\be
\label{Psiob}
\Psi (N)_{observable}= \lim \sb {L\to \infty} \frac {1}{2L} \sum \sb {n=-L}^{n=L}e^{in\theta}\Psi
(N+n)=\exp\{i(2\pi k+
\theta)N\},\quad k\in \bf Z .
\ee
In the opposite case, $p\neq 2\pi k+ \theta$,
the corresponding wave function (i.e. \it the probability amplitude\rm)
disappears: $\Psi (N)_{observable}=0$,
\it due to the complete destructive interference\rm. \par
The consequence of this topological degeneration is that the part
of  values of  momentum spectrum becomes \it unobservable \rm
in comparison with a trivial topology.\par
This fact can be treated as a\it confinement \rm  of those values which do not coincide with
\be
\label{conf}
p_k=2\pi k+ \theta, \quad 0\leq\theta\leq \pi.
\ee
The observable spectrum follows also from the constraint of 
equivalence of the points  $N$ and $N + 1$:
\be
\label{eqv}
\Psi (N)=e^{-i\theta}\Psi (N+1),\quad \Psi (N)=e^{i\pi N}.
\ee
(the $\theta$-angle is an eigenvector of the gauge transformation
$T_1 \vert \theta >=e^{i\theta}\vert \theta >$ corresponding to the
rise of the
topological number on  unit: $T_1 \vert n >=\vert n +1>$.
This theory is valid both for the
 Euclidean and Minkowski spaces).\par
As a result, we obtain the spectral decomposition of the Green function
of the free rotator (\ref{rot1})
(as the probability amplitude of the transition from the point
$ N_{in}$ to $N_{out}$) over the observable values of spectrum (\ref{conf}):
\be
\label{ampl}
G(N_{out},N_{in}|t_1)\equiv <N_{out}|\exp(-i\hat Ht_1)|N_{in}>
 = \frac{1}{2\pi}\sum\limits_{k=-\infty }^{k=+\infty }\exp\left[
 -i\frac{p_k^2}{2I}t_1+ip_k(N_{out}-N_{in})\right]~.
\ee
Using the connection with the Jacobian theta-functions \cite {Pollard}:
\be
\label{Theta}
\Theta_3(Z\vert \tau)= \sum \sb {k=-\infty} ^{k=\infty}\exp [i\pi k^2\tau+2ikZ]= 
(-i\tau)^{-1/2}\exp [\frac {Z^2}{i\pi\tau}]
\Theta_3(\frac {Z}{\tau\vert -\frac {1}{\tau}}),
\ee
we can represent the expression (\ref{ampl}) as the  sum over all the paths:
\be
\label{paths}
G (N_{out},N_{in}\vert t_1)=\sqrt{\frac {I}{4\pi i t_1}} \sum \sb {n=-\infty} ^{n=\infty}
\exp [i\theta n]\exp [+iW(N_{out},N_{in}\vert t_1)
)],
\ee
where
\be
\label{w}
W (N_{out}+n,N_{in}\vert t_1)=\frac {(N_{out}+n-N_{in})^2 I}{2t_1}
\ee
is the rotator action (\ref{rot1}).
\section { Confinement as a destructive interference.}
The topological confinement similar to the complete destructive interference
of the phase factors of the topological degeneration (i.e., \it to a
pure quantum effect \rm ) can be in the "classical non-Abelian
field theory". Recall that, at the time of the  first paper of Dirac \cite {Dir},
the so-called "classical relativistic field theories" were found in
the papers of Schr\"odinger, Fock, Klein, Weyl \cite{Fock,Weyl} as  types of
\it relativistic quantum mechanics\rm,
i.e.as  results of the primary quantization.
The phases of the gauge transformations were introduced by
Weyl \cite {Weyl} \it as pure quantum magnitudes\rm.\par
The  free rotator theory shows  that the topological degeneration can be
removed  if all the Green functions are averaged over the values of the
topological
variable and all  the possible angles of orientation of the  monopole unit
vector $\bf n$ 
in the group space (instead of the instanton averaging over
interpolations between  different vacua
in the Euclidean space).\par
The averaging over all the parameters of  degeneration can lead to the
complete destructive interference of all the colour amplitudes
\cite {Pervush3,Ilieva,Nguyen}. In this case only \it colourless \rm ("hadronic")
states \it form the complete set of physical states\rm.
Using the example
of a free rotator, we see that the disappearance of the part of
physical states due to the confinement does not
violate the composition law for a Green function:
\be
\label{comp}
G_{ij}(t_1,t_3)= \sum \sb {h} G_{ih}(t_1,t_2)G_{hj}(t_2,t_3)~,
\ee
defined as  the probability amplitude to  find the system with the
Hamiltonian $H$ in the state $j$ at the  time $t_3$  if, at the  time $t_1$,
this system was in the state $i$, where $(i; j)$ belong to the complete set
of all the states $\{h\}$:
\be
\label{compl.set}
G_{ij}(t_1,t_3)=<i\vert \exp -i \int \sb {t_1}^{t_3} H)\vert j>.
\ee
A particular case of this composition law (\ref{comp}) is unitarity of the  S-matrix:
\be
\label{un}
SS^+ = I \Longrightarrow  \sum \sb {h} <i\vert S \vert h><h \vert S^+ \vert j>= <i\vert j>,
\ee
known as the law of the probability conservation for the  S-matrix
elements ($S = I + iT$), where
\be
\label{cons}
\sum \sb {h} <i\vert T \vert h><h \vert T^* \vert j>= 2 Im<i\vert T \vert j>
\ee
(compare with (64.2), (71.2) in \cite {BLP}).
The left-hand side of this law is similar to the spectral series of the
free rotator (\ref{ampl}).\par
The destructive interference keeps only the colourless "hadronic" states.
 Whereas the right-hand side of this law, far from resonances, can
be represented by the perturbation series over the Feynman diagrams
that follow from the Hamiltonian. Due to the gauge invariance,
$H[A^{(n)},q^{(n)}]=H[A^{(0)},q^{(0)}]$,
where $q$ are the fermion (quark) degrees
of freedom.
This means that the Hamiltonian $H[A^{(0)},q^{(0)}]$ depends on the
Gribov phase in its BPS monopole form (\ref{phase})(the Gribov phase (\ref{phase})
\it is a colour scalar\rm ), but it does not depends on the
Gribov phase factors (\ref{mon.deg}).
The considered above holonomy 
theory (\ref{bg})-(\ref{coh}) allow us to draw the conclusion that \it the colour
confinement in the considered YM theory, with the Gribov equation \rm (\ref{Gribov.eq})
\it and its vacuum solution \rm (\ref{phase}), is determined by the restricted
holonomy group \rm $\Phi^0$ \it generated by the  zero topological 
sector of this YM theory\rm (more precisely, by the YM fields of this
sector satisfied the Coulomb gauge (\ref {Aparallel})).
We can interpret this as a \bf confinement
criterion in QCD \rm (which is also true for the gluonic theory with the
$SU(3)_{col}\to SU(2)$ spontaneous break- down). \par
 Thus, Hamiltonian $H$  contains the perturbation series in terms only of
zero degree of the map fields  (i.e., in terms of  \it constituent colour
particles \rm ) that can be identified \it with  Feynman partons\rm.
The Feynman path integral as the generating
functional of this perturbation series is an analogue of the sum
over all the paths of the free
rotator (\ref{paths}).\par
Therefore, confinement, in the spirit of the complete destructive
interference of  colour amplitudes \cite {Pervush2,Pervush3,Nguyen},
and the law of the probability conservation for the S-matrix elements, (\ref{cons}),
lead to the \it Feynman quark-hadronic duality \rm  that is the base of all the
partonic models \cite{Feynman} and the QCD applications \cite{Efremov}.
 The quark-hadronic duality gives a method of  the direct experimental measurement
 of  quark and gluon quantum numbers from a deep-
inelastic scattering cross-section \cite {Feynman}.
For example, according to Particle Data Group,
the ratio of  the sum of the probabilities of the $\tau$-
decay hadronic  modes to the probability of the $\tau$-decay muonic  mode is
\be
\label{ratio}
\frac {\sum \sb {h} w_{\tau \to h}}{w_{\tau \to \mu}}=3.3 \pm 0.3 .
\ee
This is the left-hand side of the Eq.(\ref{cons})
normalized to the value of the leptonic  mode probability of the $\tau$-decay.
On the right-hand side of the Eq.(\ref{cons}) we have the ratio of the imaginary part
of  the sum over  the quark-gluonic diagrams (in terms of constituent
fields free from any Gribov phase factor) to the one of the leptonic
diagrams. In the lowest order of QCD perturbation on the right-hand
side we get the number of colours $N_c$, therefore
\be
\label{Nc}
3.3 \pm 0,3 = N_c.
\ee
Thus, the degeneration of initial data  can explain us not only
"\it why we do not see  quarks\rm",
but also "\it why we can measure their
quantum numbers \rm ".
This mechanism of confinement, due to a quantum interference of the
phase factors of the topological degeneration,
 disappears after a change of the
"physical" sources:  $A^*J^*\Longrightarrow A J$,
called the transition to  another gauge in the gauge-fixing method.
Then, for example, the Coulomb gauge (\ref {Aparallel}) is not valid.
The Gribov ambiguity equation (\ref{Gribov.eq}), which describes the
ambiguity in the choice of the YM fields satisfied 
the Coulomb gauge (\ref {Aparallel}) (having the Gribov
phase (\ref{phase}) as its solution), turns then in some formal 
differential equation, without of any physical sense. The restricted
holonomy group $\Phi^0$, constructed on the transverse YM fields satisfied
the Coulomb gauge (\ref {Aparallel}), becomes trivial in this case.
This means, in turn, that the confinement criterion, considered above,
is not valid also.\par
Instead of the hadronization and confinement, we obtain then
the scattering amplitudes  of the free partons only. But these 
amplitudes do not exist as physical observable in the Dirac
quantization scheme, which depends on initial data.
\section { The U(1)-problem.}
 The  value of the vacuum chromo-magnetic field $<B^2>$ can be estimated
 by the description of a process with an anomaly.
 The simplest process of such  type is the
 of a pseudo-scalar bound state with an anomaly.
 In gauge theories there is the universal effective  action
 for the description of  this interaction:
\be
\label{inter}
 W_{eff}=\int dt \left\{\frac 1 2 \left({\dot\eta_M}^2-M_P^2{\eta_M}^2\right)
 V +
 C_M\eta_P \dot X[A^{(N)}] \right\}~,
\ee
 where $\eta_M$ is a bound state with the mass $M_P$ in its rest frame of reference, and
 $X[A^{(N)}]$ is the topological  "winding number" functional.
 In 3-dimensional QED${}_{(3+1)}$ this action,  with the constant \cite {Pervush2}
 \be
\label{C}
 C_M=C_{ positronium}=\frac{\sqrt{2}}{m_e}8{\pi}^2
 \left(\frac{\underline{\psi}_{Sch}(0)}{m_e^{3/2}}\right)~,
 \ee
 describes the decay of a positronium $\eta_M=\eta_P$ into two photons
 that are in the  "winding number" functional
 \be
\label{wn}
 \dot X_{ QED}[A]= \frac{e^2}{16 \pi^2}
 \int d^3x F_{\mu \nu}{}^*F^{\mu \nu}\equiv\frac{e^2}{8 \pi^2}
 \int d^3x \varepsilon_{ijk}\dot A^i(\partial^jA^k- \partial^kA^j)~.
 \ee
 In 1-dimensional QED${}_{(1+1)}$ this action (\ref{inter}), with the constant
 $C_M=2\sqrt{\pi}$ and the "winding number" functional
 \be
\label{funct}
 \dot X_{ QED}(A^{(N)})=\frac{e}{4\pi}
 \int\limits_{-V/2 }^{V/2 }dx F_{\mu\nu}
 \epsilon^{\mu\nu}= \dot N(t)~\Rightarrow~~F_{01}=\frac{2\pi\dot N}{eV}~,
  \ee
 describes the  mass of the Schwinger
 bound state $\eta_P=\eta_{ Sch.}$ if the action (\ref{inter})
 is added by the action of the Coleman electric field \cite{Ilieva,Gogilidze}:
 \be
\label{el.f}
  W_{ QED} =\frac{1}{2}\int dt  \int\limits_{-V/2 }^{V/2 }dx F^2_{01}=
 \int dt\frac{\dot N^2 I_{ QED}}{2}~,
 \ee
 where
 \be
\label{IQED}
 I_{ QED}=\left(\frac{2\pi}{e}\right)^2\frac{1}{V}~.
  \ee
 It is easy to see that the diagonalization of the total Lagrangian
 of the 
 \be
\label{Lag}
 L=[\frac{\dot N^2I}{2}+C_M\eta_M \dot N] =
 [\frac{(\dot N+C_M\eta_M/I)^2I}{2}- \frac{C_M^2}{2IV} \eta_M^2V ]
 \ee
 type leads to the mass of a pseudo-scalar meson in
  QED${}_{(1+1)}$:
\be
\label{mQED} 
  \triangle {M}^2=\frac{C_{M}^2}{I V}=\frac{e^2}{\pi}~.
\ee
 In QCD${}_{(3+1)}$ a similar action for a pseudo-scalar meson
 $\eta_M=\eta_0$ was proposed in \cite{Veneziano}, where the "winding number"
 functional was given by
 \be
\label{wi}
 \dot X_{ QCD}[A^{(N)}]= \frac{g^2}{16 \pi^2}
  \int d^3x F^a_{\mu \nu}{}^*F_a^{\mu \nu}=  \dot N(t)+\dot X[A^{(0)}]
 \ee
 and
 \be
\label{Ceta}
 C_M=C_{\eta}=\frac{N_f}{F_\pi}\sqrt{\frac 2 \pi } , ~~~(N_f=3)~.
 \ee
 As we have seen,  QCD${}_{(3+1)} $ has the chromo-electric monopole:
 \be
\label{chm}
 F^a_{0i}=\dot N D_i^{ab}({\Phi})\Phi_{b0} = \dot N B_i^a({\Phi})
 \frac{2\pi}{\alpha_s V<B^2>}~,
  \ee
 with the normalization
 \be
\label{norm}
 \frac{g^2}{8 \pi^2}\int d^3D_i^{ab}({\Phi})\Phi_{b0} B^i_a({\Phi})=1~.
  \ee
 The action (\ref{inter}) should be added by the action of the
 topological dynamics of the zero mode $\dot N$:
 \be
\label{t.d}
  W_{ QCD} =\frac{1}{2}\int dt  \int\limits_{V }^{ }d^3x F^2_{0i}=
 \int dt \frac{\dot N^2 I_{ QCD}}{2}~,
  \ee
 with the mass $ I_{ QCD}$ determined by the vacuum magnetic field:
 \be
\label{mQCD}
 I_{ QCD}=\left(\frac{2\pi}{\alpha_s}\right)^2\frac{1}{V<B^2>}~.
 \ee 
 In  QCD${}_{(3+1)}$ the equation for the diagonalization,
  (\ref{Lag}), leads to an additional mass of the
   $\eta_0$ meson:
 $$
  L_{eff}= \frac{1}{2}[{\dot\eta_0}^2-\eta_0^2(t)(m_0^2
 +\triangle m_{\eta}^2)]V~,
 $$
 $$
  \triangle {m_\eta}^2=\frac{C_{\eta}^2}{I_{ QCD}V}
 = \frac{N_f^2}{F_{\pi}^2}\frac{\alpha_s^2<B^2>}{2\pi^3}.
 $$
 This result allows us to estimate the value of
  the vacuum chromomagnetic field in QCD${}_{(3+1)}$:
\be
\label{Bav} 
 <B^2>=\frac{}{}\frac{2\pi^3F_{\pi}^2\triangle
 {m_\eta}^2}{N_f^2\alpha_s^2}=\frac{0.06 GeV^4}{\alpha_s^2}
 \ee 
 (see also \cite{David3}). After the calculation we can remove
  the infrared regularization $V \to \infty $.
\section {Conclusion.}
 The main problems of the discussion of stable
 vacuum states in some non-Abelian theory are the classes of functions
 and singularities.
 These problems exist in all models of the QCD vacuum, including
 instantons described by the $\delta$-type singularities in the
 Euclidean space.\par
 Mysteries of the nature are not only the actions and symmetries, but
 also the class of functions with  finite energy densities used
 in quantum field theories (including QED)
 for the description of physical processes.
 If we explain any effect by these  singularities,
 choosing a model of the nontrivial QCD vacuum, we should answer the
 questions: "Where are singularities of this vacuum from?" and
 "What is a physical origin of these singularities?".
 \par
 We  presented  here the model of the vacuum in the  Yang-Mills (YM)
 theory in the monopole class of functions with the finite energy density
 without any singularity in a finite volume,
 as  a consequence ("smile") of the scalar Higgs field that
 disappears (like the Cheshire cat) from the
 spectrum of physical excitations of the theory in the limit of the
 infinite spatial volume. In other words,
 we proved that there exists a mathematically correct model of the YM vacuum, with the
 finite physical energy-momentum spectrum in the Minkowski space,
 constructed from the well-known Bose condensate of the Higgs scalar field
 in the limit of its infinite mass.\par
 The  $SU(2)$ symmetry of the
 YM vacuum is broken down spontaneously. The break-down
 $SU(2)\to U(1)$ is
 realized in the presence of  the Higgs $SU(2)$ isovector.
 If the Higgs field goes to the statistical (vacuum) expectation value
 at the spatial infinity, this leads to the nontrivial
 topological structure of the remaining group of symmetry, $U(1)$,
 induced by the Higgs vacuum expectation value.
  This nontrivial topological structure means
 the presence of  topological (magnetic) charges in this theory, i.e.
 \it the inevitability of the monopole configurations of the YM vacuum with
 finite energies\rm.\par
 We have considered our theory in the BPS limit  when the self-interaction
between the Higgs particles goes to zero. This allows us to consider the
Higgs particles (in the limit of  their infinite number, i.e. at the
level of  statistical physics) as \it an ideal gas\rm.
We imposed an additional condition of stationarity of
this ideal gas (Bose condensate).
This choice influences the stationary nature of the monopole
configurations of the YM vacuum. \par
The Bogomol'nyi equation obtained issuing from the
evaluation  of the  lowest bound of energy for the   monopole solutions
(the latter one depends on the vacuum expectation value $m/\sqrt{\lambda}$)
allowed us to find
the monopole configurations of the YM vacuum as
Bogomol'nyi-Prasad-Sommerfeld (BPS) or Wu-Yang  monopoles
(obtained as  infinite volume limits of  BPS monopoles).\par
We described
the topological degeneration  of initial data for  monopole solutions  at
nonzero values of the  topological charge. This topological degeneration
manifests itself as Gribov copies of the covariant Coulomb gauge considered
 as  zero initial data for the Gauss law constraint.
These Gribov copies are defined by a
solution of the Gribov ambiguity  equation (\ref{Gribov.eq})  in the class of functions of the BPS monopole
type
for the Higgs vacuum field.
\par
The  Gribov equation (\ref{Gribov.eq}) describes the correct cohomological
structure of the YM vacuum at the spatial infinity.
There exists an one-to-one
correspondence between the set of cohomology classes of YM
fields and the set of  Gribov copies of the Coulomb gauge (\ref {Aparallel}).
This cohomological structure corresponds to
the elements of the holonomy group $H$ constructed on the transverse YM
fields. The unit element of the holonomy group $H$ is degenerated
with respect to the class of exact 1- forms (with the zero
topological charge) induced by the Coulomb gauge (\ref {Aparallel}) 
and the Bogomol'nyi equation (\ref{Bog}). \par
 The Yang-Mills fields  are considered as  sums of
vacuum   fields (monopoles) and  weak perturbation excitations over this vacuum
(multipoles).
We suppose that these excitations have the same topological numbers
as the vacuum components. \par
The important point of our investigations is
that the  square of the  Gibbs expectation
value of the magnetic tension, $<B^2>$, is different from zero.
This is a direct
analogy  of the \it Meisner effect \rm in a superconductor.
In the language of the group theory it means the spontaneous
 break-down of  the $U(1)$ symmetry.
\par
We proved that there is the continuous topological variable $N(t)$
defining the zero mode of the Gauss equation and depending on
the time $t$; it plays the
role of the non-integer degree of the map.
The calculations led to the action of a free rotator
with the rotation momentum $I$ depending  on $<B^2>$ and
\it  real spectrum of momentum\rm. This
spectrum  describes the rotation of the  Yang-Mills vacuum
(as a Bose condensate, depending on the vacuum expectation value $<B^2>$ )
as a whole system.
\par
The considered  nontrivial topological structure of the vacuum in the
 YM theory  can be in  other non-Abelian theories.
For example,
 there is the spontaneous $SU(3)_{col}\to SU(2)$
break-down with the antisymmetric choice of the Gell-Mann matrices
$\lambda_2, \lambda_5, \lambda_7$, which
 leads to the Wu-Yang monopole (see the
formulas (3.24-3.25) in \cite {David2}).
The essential point of the theory \cite {David2} was \it the mix of
  world and  group indices \rm in the
construction of the   Wu-Yang monopole\rm.
One can consider the behaviour of
quarks  in the Wu-Yang monopole field and to write down the Green
function of  a quark (see (4.9)-(4.13) in \cite {David2}). \par
Physical arguments in the favour  of the considered theory of
the physical vacuum
are an additional mass of the $\eta '$-meson in QCD,
the rising potential and topological confinement \cite{Pervush2}.\par
We calculated explicitly the hadronization potential
$V_1$ as one of the components in the decomposition of the Green
function of the Gauss (Gribov )equation in the presence of the Wu-Yang monopole
by the complete set of  orthogonal vectors in  the colour space.
This is a non-local monopole effect.\par
We proved that  the topological
confinement can lead to the colour confinement in QCD in the form of
the complete destructive interference of the phase factors
of the topological degeneration.
This means that
only the colourless ("hadronic") states can be treated as  physical states.
\par
The Hamiltonian of QCD depends only on the Gribov phase (\ref{phase})
as a colour isoscalar.
As a result, the criterion of the colour confinement in QCD, 
in the  Coulomb  gauge (\ref{Aparallel}),
is the 
existence of the nontrivial restricted holonomy group $\Phi ^0$ 
constructed on the transverse YM
fields of the zero topological sector.\par
The Lorentz covariance can be carried out by the Lorentz rotation of the
time axis $l_\mu^{(0)}$ along the complete momentum of each of the
physical states, i.e. by the transition to the frame of reference
where the initial data and the spectrum of these states are measured
\cite {David2}.
\par
All these "smiles" of the Higgs scalar field disappear if we replace
the fundamental Dirac variables \cite{Dir,Nguyen2,FJ} and change
the gauge of their physical sources in order to obtain
the conventional Faddeev-Popov integral \cite{Fadd3} as a realization of
the Feynman heuristic quantization \cite{AGP}.
This change removes all the time axes of the physical states, all the initial
data, with their degeneration and destructive interference, and all
monopole effects, including instantaneous interactions forming
non-local bound states of the types of atoms in QED or hadrons in QCD.
In other words, the "smiles" of the Higgs field show us
the limitedness of the Faddeev-Popov heuristic path integral.
 The generalization of the
Faddeev theorem of equivalence \cite{Fadd1} (that is valid for  local scattering
processes) to the region of  non-local processes removes
both the initial data
and the Laplace possibility of explaining (by these data) the non-local physical
effects of the type of hadronization and confinement  in
this world.
\section*{Acknowledgments.}
 We are  grateful to Prof. B .M.~Barbashov, E.  A.  Kuraev  and A.  A.  Gusev for fruitful discussions,
 Prof. A.  S.  Schwarz for his critical remark
 and Prof. C. K.  Zachos who kindly informed us about an
 infinite Higgs mass coupling  analysis.


\begin{thebibliography}{90}
\bibitem{Mat}  S. G. Matinyan, G. K. Savidy, Nucl. Phys. \bf B 134\rm, 539 (1978).
\bibitem {Gr} D. J. Gross, F. Wilczek, Phys. Rev. \bf D 8\rm, 3633 ( 1973);\\
 H. D. Politzer, Phys. Rev. \bf D 8\rm, 3636 ( 1973);\\
 A. A. Vladimirov, D. V. Shirkov, Soviet.
 Phys. Usp. \bf 22\rm , 860 (1979).
\bibitem {Bel} A. A. Belavin, et al., Phys. Lett. \bf 59\rm, 85 (1975);\\
 R. Jackiw , C. Rebbi, Phys. Lett. \bf B 63\rm, 172 (1976);\\
 C. G. Jr. Callan, R. F. Dashen ,D. J. Gross,  Phys.Rev. \bf  D 17\rm, 2717 ( 1977).
\bibitem {Al.S.}A. S. Schwarz,  Kvantovaja Teorija Polja i Topologija, 1st edn.(Nauka, Moscow 1989).
\bibitem {N.N.}N. N. Bogoliubov, J. Phys. \bf 9\rm, 23 (1947); \\ N. N. Bogoliubov,
 V. V. Tolmachev , D. V. Shirkov,   Novij Metod v Teorii
Sverchprovodimosti, 1st edn.(
Izd-vo AN SSSR 1958:p.p.5-9).
\bibitem {Logunov}  N. N. Bogoliubov, A. A. Logunov, A. I. Oksak, I. T. Todorov, Obshie Prinzipi Kvantovoj 
Teorii Polja, 1st edn. ( Nauka, Moscow 1987).
\bibitem {Leonid} L. D. Lantsman, The one Example of  Lorentz Group[ hep-th /0104182].
\bibitem {Kolm} A. N. Kolmogorov, S. V. Fomin,  Elementi Teorii Funktsij i Funktsionalnogo Analisa, 1st edn
(Nauka, Moscow 1976).
\bibitem {Ryder}L. H. Ryder, Quantum Field Theory, 1st edn. ( Cambridge University Press 1984).
\bibitem {Cheng} T. P. Cheng, L.- F. Li, Gauge Theory of Elementary Particle Physics, 3rd edn.
(Oxford University 
Press 1988).
\bibitem{Linde}
 A. D. Linde, Elementary Particle Physics and Inflationary Cosmology, 1st edn.
(Nauka, Moscow 1990).
\bibitem{Landau3} L. D. Landau,  E. M. Lifschitz,  Theoretical Physics, v. 3. Quantum Mechanics, edited by 
L. P. Pitaevskii, 4th edn.
(Nauka, Moscow 1989).
\bibitem{Pervush1}
 V. N. Pervushin, Teor. Mat. Fiz. \bf 45\rm, 327 (1980);\\
English translation in Theor. Math. Phys. \bf 45 \rm, 1100 (1981).
\bibitem{Hooft}G. 't Hooft, Nucl.Phys. \bf B 79\rm, 276 (1974).
\bibitem{Polyakov} A. M. Polyakov, Pisma  GETF \bf 20\rm, 247 (1974).
\bibitem{BPS} M. K. Prasad, C. M. Sommerfeld, Phys. Rev. Lett. \bf 35\rm, 760 (1975);\\
 E. B. Bogomol'nyi, Yad. Fiz. \bf 24\rm, 449 (1976).
\bibitem{Gold} R. Akhoury,  Ju-Hw. Jung, A. S. Goldhaber, Phys. Rew. \bf 21\rm, 454 (1980).
\bibitem{Switz} R. M. Switzer, Algebraic Topology-Homotopy and Homology, 1st edn.( Springer Verlag, 
Berlin, Heidelberg, New York 1975).
\bibitem{Pervush2} V. N. Pervushin, Dirac Variables in Gauge Theories [ hep- th/ 0109218]. 
\bibitem{Vlad} V. S. Vladimirov, Yravnenija Matematicheskoj Fiziki, 5th edn. (Moscow, Nauka 1988).
\bibitem{Schwinger}  J. Schwinger, Phys. Rev. \bf 127\rm, 324 (1962).
\bibitem{Fadd1} L. D. Faddeev, Teor. Mat. Fiz. \bf 1\rm, 3 (1969),  in Russian.
\bibitem{Gitman} D. M. Gitman, I. V. Tyutin,  Kanonicheskoje Kvantovanije Polej
so Svjasjami, 1st edn. (Nauka, Moscow 1986).
\bibitem{Slavnov} L. D. Faddeev, A. A. Slavnov, 
Introduction to Quantum Theory of Gauge Fields, 2nd edn. (Nauka, Moscow 1988). 
\bibitem{Fadd2} L. D. Faddeev, A. A. Slavnov,  Gauge  Fields: Introduction to Quantum Theory, 1st edn. 
(Benjamin-Gummings 1984).
\bibitem{Fradkin1} E. S. Fradkin, I. V. Tyutin,  Phys. Rev. \bf D  2\rm, 2841 (1970).
\bibitem{Nils} N. K. Nielsen, P. Olesen,  Nucl. Phys. \bf B 144\rm, 376 (1978).
\bibitem{Gribov} V. N. Gribov,  Nucl. Phys. \bf B  139\rm, 1 ( 1978).
\bibitem{Pervush3} V. N. Pervushin,  Riv. Nuovo Cim. \bf  8 \rm, N \bf 10\rm, 1 (1985).
\bibitem{Ilieva} N. Ilieva, V. N. Pervushin,  Sov. J. Part. Nucl. \bf 22\rm, 573 (1991).
\bibitem{Nguyen}  V. N. Pervushin, Nguyen Suan Han, Can. J. Phys. \bf 69\rm, 684 (1991).
\bibitem{Arsen}A. M. Khvedelidze, V. N. Pervushin,  Helv. Phys. Acta  \bf 67\rm, 637 (1994).
\bibitem{Jack} L. D. Faddeev,  in Proceeding of the 4 Int. Symposium on non-local Quantum 
Field Theory, Dubna, 1976 (JINR D1-9768),  p. 267;\\
R. Jackiw, Rev. Mod. Phys. \bf 49\rm, 681 (1977).
\bibitem{Niemi}L. D. Faddeev, A. J. Niemi,  Nature \bf 387\rm, 58 (1997).
\bibitem{David1} D. Blaschke, V. N. Pervushin, G. R$\ddot o$pke,  
in Proceeding  of the Int. Seminar Physical variables 
in Gauge Theories, Dubna, September 21-25, 1999,
edited by A. M. Khvedelidze, M. Lavelle, D. McMullan, and V. Pervushin
(E2-2000-172, Dubna  2000), p. 49; 
[hep-th/0006249].
\bibitem{Bogolubskaja}
 A. A. Bogolubskaya, Yu. L. Kalinovsky, W. Kallies, V. N. Pervushin,
 Acta Phys. Polonica \bf 21\rm, 139 (1990).
\bibitem{Yura} V.N. Pervushin, Yu. L. Kalinovsky, W. Kallies, N. A. Sarikov,
 Fortschr. Phys. \bf 38\rm, 333 (1990).
\bibitem{Wu} T. T. Wu, C. N. Yang, Phys. Rev. \bf D  12\rm, 3845 (1975).
\bibitem{David2}D. Blaschke, V. N. Pervushin, G. R$\ddot o$pke,
\rm Topological Gauge Invariant Variables in QCD [ hep-th /9909193].
\bibitem{A.I.}
{\normalsize  A. I. Achieser, V. B. Berestetskii, Quantum  Electrodynamics, 3rd edn. (Moscow, Nauka 1969).}
\bibitem{Fadd3} L. Faddeev, V. Popov, Phys. Lett. \bf B  25\rm, 29 (1967).
\bibitem{Dir}P. A. M. Dirac, Proc. Roy. Soc. \bf A 114\rm, 243 (1927); Can. J. Phys. \bf 33\rm, 650 (1955).
\bibitem{Postn3} M. M. Postnikov,  Lektsii po Geometrii
(Semestr 3, Gladkie Mnogoobrazija), 1st edn.  (Nauka, Moscow 1987).
\bibitem{Postn4}M. M. Postnikov,  Lektsii po Geometrii
(Semestr 4, Differentsialnaja  Geometrija), 1st edn. ( Nauka, Moscow 1988).
\bibitem{Coleman}S. Coleman,  Ann. Phys. (N. Y.) \bf 93\rm, 267 (1975); ibid. \bf 101\rm, 239 (1976).
\bibitem{Gogilidze}S. Gogilidze, N. Ilieva, V. Pervushin,  Int. J. Mod. Phys \bf A 14\rm, 3531 (1999).
\bibitem{Kamke} 
 E.Kamke, Differentialgleichungen Losungsmethoden und 
Losungen (v. 1. Gewohnliche Differentialgleichungen), 6th edn. (Leipzig 1959).
\bibitem{Werner}V. N. Pervushin, Yu. L. Kalinovsky, W. Kallies, N. A. Sarikov,
 Fortschr. Phys. \bf 38\rm, 333 (1990).
\bibitem{Pollard}  B. R. Pollard,  An Introduction to Algebraic Topology, Notes of Lectures Given During The 
Session 1976-1977, 1st edn. (University of
Bristol).
\bibitem{Fock} V. Fock,  Z. Phys. \bf 39\rm, 226 ( 1926); ibid. \bf 57\rm, 261 (1929).
\bibitem{Weyl} H. Weyl,  Z. Phys. \bf 56\rm, 330 (1929).
\bibitem{BLP}L. D. Landau, E. M. Lifshitz, 
Theoretical Physics, v. 4. Quantum Electrodynamics (V. B. Berestetskii,
E. M. Lifshitz, L. P. Pitaevskii), edited by L. P. Pitaevskii, 3rd edn. 
(Nauka, Moscow 1989).
\bibitem{Feynman} R. P. Feynman,  Photon Hadron Interaction, 1st edn.( New-York, N.Y. 1972).
\bibitem{Efremov}A. V. Efremov, A. V. Radyushkin,
 Riv.Nuovo Cimento \bf 3\rm, N. \bf 2\rm, 1  (1980). 
\bibitem{Veneziano}{ G. Veneziano},  {Nucl. Phys. \bf B 195}\rm, 213 ( 1979).
\bibitem{David3}{ D. Blaschke, et al.},  {Phys. Lett. \bf B  397}\rm, 129 (1997). 
\bibitem{Nguyen2}{ Nguyen Suan Han, V. N. Pervushin},  Mod. Phys. Lett. \bf A  2\rm, 367 (1987).
\bibitem{FJ} L. D.  Faddeev, R. Jackiw, Phys. Rev. Lett. \bf 60\rm, 1692 ( 1988).
\bibitem{AGP}S. A. Gogilidze, A. M. Khvedelidze, V. N. Pervushin, 
J. Math. Phys. \bf 37\rm, 1760 (1996);
 Phys. Rev. D \bf 53\rm, 2160  (1996);  Phys.
 Particles and Nuclei \bf 30\rm, 66 (1999).
\end{thebibliography}
\enddocument